\documentclass[11pt,letterpaper]{article}

\usepackage{graphicx}   
\usepackage{lscape}
\usepackage{array,multirow,bigdelim}
\usepackage{xcolor}
\usepackage{ulem}
\usepackage{hyperref}
\hypersetup{colorlinks,urlcolor=black,citecolor=black,linkcolor=black,filecolor=black}
\usepackage{amsmath}

\advance \topmargin by -\headheight
\advance \topmargin by -\headsep

\evensidemargin \oddsidemargin
\begin{document}
	
\begin{titlepage}
\vspace*{-0.5in}
		
	\begin{center}
			{\Large {\bf Recent Advances in Semiclassical Methods Inspired \\
					by Supersymmetric Quantum Mechanics}}
	\end{center}
		\vskip .2in
		
	\begin{center}
			{Asim Gangopadhyaya\footnote{agangop@luc.edu}, 
			 Jonathan Bougie\footnote{jbougie@luc.edu}, and
			 Constantin Rasinariu\footnote{crasinariu@luc.edu}}
			
			\par 
			\vskip .1in 
			\noindent
			Physics Department,\\ Loyola University Chicago, \\1032 W Sheridan Rd.
			Chicago, IL 60660
		\end{center}
		
	\begin{abstract}
		Semiclassical methods are essential in analyzing quantum mechanical systems. Although they generally produce approximate results, relatively rare potentials exist for which these methods are exact. Such intriguing potentials serve as crucial test cases for semiclassical approximations. Recent research has demonstrated a deep connection between  supersymmetric quantum mechanics and the exactness of semiclassical methods. Specifically, the mathematical form of conventional shape-invariant potentials guarantees exactness in several related situations. In this manuscript, we review these recent results and discuss their significance.
		
	\end{abstract}
\setcounter{tocdepth}{1}
\tableofcontents

\end{titlepage}


\section{Introduction}	

The WKB or JWKB semiclassical approximation method developed by Jeffreys, Wentzel, Kramers, and Brillouin \cite{Jeffreys1924,Wentzel1926,Kramers1926,Brillouin1926,Dunham1932} remains relevant in quantum mechanics and other areas of physics even today, approximately 100 years after its invention. A recent series of papers \cite{Bougie2018, Gangopadhyaya2020, Gangopadhyaya2021, Gangopadhyaya2023} has uncovered new relationships between the WKB method, the related supersymmetric WKB method (SWKB), and the framework of supersymmetric quantum mechanics (SUSYQM). In this review paper, we discuss recent advances in this field.

We will first provide a brief overview of these semiclassical methods, the SUSYQM formalism, and shape invariance in Sec.~\ref{sec:Background}. We then use the properties of SUSYQM to categorize shape-invariant potentials (SIPs) and their connections in Sec.~\ref{sec:SIPs}. We then explore the following connected discoveries:
\begin{itemize}
	\item The SWKB method has long been known to produce exact eigenenergies for an entire class of potentials known as conventional shape-invariant potentials \cite{Comtet1985,Dutt1986,Eckhardt1986,Raghunathan1987,Dutt1991,Inomata1994}. Recent work\cite{Gangopadhyaya2020}  proved that this exactness can be derived from their mathematical structure for the case of unbroken supersymmetry (SUSY). We discuss this work in Sec.~\ref{sec:UBSWKB}.
	
	\item In the case of broken SUSY, \cite{Inomata1994, Inomata1993a, Inomata1993b, Junker2019,Eckhardt1986} a modified version of SWKB (BSWKB) applies. In  Sec.~\ref{sec:BSWKB} we present a recent result \cite{Gangopadhyaya2021} proving the exactness of BSWKB for conventional SIPs.
	
	\item The WKB method is not exact for all conventional SIPs; however, it generates exact energies for the specific cases of the  one-dimensional harmonic oscillator and Morse potentials \cite{Weiner1978a, Weiner1978b, Pippard1983, Nieto1985, Poppov1991a, Poppov1991b, Mur1991, Dalarsson1993, Maitra1996, Chebotarev1998, Park1998, Price1998, Friedberg2001, Jaffe2010}. Adding a term to the potential, known as the Langer correction (although it was first discovered by Kramers \cite{Kramers1926}), produces exact WKB eigenvalues for the Coulomb and 3D oscillator potentials, and also corrects the behavior of wavefunctions near singular points\cite{Kramers1926, Langer1937, Langer1949, Bailey1964, Froman1965}. A recent paper \cite{Gangopadhyaya2023} proved that a generalized Langer-like correction makes all conventional SIPs WKB-exact and regularizes their wavefunctions near singularities. This exactness is intrinsically linked with the SWKB-exactness of these potentials. We review these results in Sec.~\ref{sec:langer}.
	
	\item Extended SIPs are closely related to the conventional SIPs, in that each known extended SIP is isospectral with a conventional SIP that serves as its kernel \cite{Quesne2008,Quesne2009,Odake2009, Odake2010, Odake2011, Odake2013, Tanaka2010,Bougie2010, Bougie2012}. 	In Sec.~\ref{sec.Extended}, we discuss a paper \cite{Bougie2018} demonstrating that SKWB exactness is not guaranteed for extended SIPs, despite their additive shape invariance. 
\end{itemize}

We summarize these results in Sec.~\ref{sec:conclusions}.

\section{Background: Semiclassical Methods, SUSYQM, and Shape Invariance}\label{sec:Background}

In this section, we discuss important background to the recent advances addressed in this review, including semiclassical methods (Sec.~\ref{sec:semiclassical}), SUSYQM (Sec.~\ref{sec:susyqm}), and shape invariance (Sec.~\ref{sec.shapeinvariance}).

\subsection{Semiclassical WKB, SWKB, and BSWKB Approximations}\label{sec:semiclassical}
We consider a quantum system to be exactly solvable when its eigenvalues can be analytically expressed in terms of the system's parameters \cite{Cooper2001, Gangopadhyaya2017}. 
Since this is not true of most potentials, solving the Schrödinger equation often requires approximation methods. While perturbation theory is typically employed for systems that are ``close" to being exactly solvable, the WKB approximation offers a non-perturbative approach to approximating eigenenergies and wavefunctions by assuming a potential $V(x)$ with slow spatial variation in $x$. In this paper, we focus on the eigenvalues $E_n$ of the system, which are approximated using the WKB quantization condition:
\begin{equation}
	\int_{x_L}^{x_R} \sqrt{E_n-V(x)} ~dx = \left( n+\nu \right) \pi \hbar~, \label{eq.Bohr-Sommerfeld}
\end{equation}
where $\nu$ is a fractional number known as the Maslov index, which depends on the specific properties of the potential. The integration limits ${x_L}$ and ${x_R}$ are the classical turning points where  $V(x)$ equals $E_n$. 

The quantization condition from Eq.~(\ref{eq.Bohr-Sommerfeld}) generally provides good approximations for large values of $n$. However, for certain smooth potentials, its validity can extend \cite{Karnakov2013} to $n\sim 1$. Remarkably, this usually approximate condition yields exact eigenvalues for both the one-dimensional harmonic oscillator and Morse potentials with $\nu = \frac12$.

In 1926, Kramers\cite{Kramers1926} discovered that the WKB method was inaccurate for the Coulomb potential's radial wave function near the origin; however,  adding an {\it ad hoc} term $\frac{\hbar^2}{4r^2}$ to $V(r)$ leads to the correct behavior of the wave function near the singularity.  This adjustment also makes the quantization condition exact for the eigenergies of this system \cite{Young1930}. Langer later showed that the same correction made the radial oscillator energies exact, and argued that the correction was needed due to the semi-infinite domain of Coulomb and the radial oscillator; this correction became known as the Langer correction. \cite{Langer1937, Langer1949, Bailey1964, Froman1965}

SUSYQM relates ``partner potentials'' through the use of a superpotential $W(x,a)$, as we will discuss in Sec.~\ref{sec:susyqm}.   In this context, a modified version of WKB (known as SWKB) applies to the superpotential  \cite{Comtet1985,Eckhardt1986,Dutt1986,Dutt1991,Inomata1994,Gangopadhyaya2020}, which states that
\begin{equation}\label{eq:swkb}
	I_S \equiv 
	\int_{x_L}^{x_R} \sqrt{E_n(a)-W^2(x,a)} ~dx = n \hbar \pi~,
	\quad \mbox{where}~ n = 1,2,3,\cdots~,
\end{equation}
where the integration limits are given by the zeros of the integrand. In the case of unbroken supersymmetry (which we will discuss in Sec. \ref{sec:susyqm}), this SWKB condition is exact for an entire class of potentials, known as conventional shape-invariant potentials, without the need for a Langer-type correction \cite{Adhikari1988}.

While the above condition applies for unbroken SUSY, Inomata and Junker \cite{Inomata1994, Inomata1993a, Inomata1993b, Junker2019}, and Eckhardt \cite{Eckhardt1986} independently proposed a modified SWKB condition for systems with broken supersymmetry, which we will call the BSWKB condition:
\begin{eqnarray}
	I_S \equiv 
	\int_{x_L}^{x_R} \sqrt{E_n(a)-W^2(x,a)} ~dx  = \left( n+\frac12\right)  \pi\hbar~, \quad \mbox{where}~ n = 0,1,2,\cdots \label{eq:bswkb}~.
\end{eqnarray}
This condition is exact for all conventional SIPs with bound states in the broken-SUSY phase \cite{Gangopadhyaya2021}.

\subsection{SUSYQM}\label{sec:susyqm}Supersymmetric quantum mechanics is a generalization of the algebraic ``ladder'' method for the harmonic oscillator developed by Dirac and F\"ock.  In SUSYQM, a Hamiltonian is written as the product of two linear differential operators: $A^{\pm} \equiv \mp\, \hbar \; d/dx + W(x,a)$, which are Hermitian conjugates of each other. The function $W(x,a)$ is called the superpotential. These operators generate two ``partner" Hamiltonians, $H_{\mp}$:
\begin{equation}\label{eq.LadderOperators}
	A^{\pm}A^{\mp}~=~H_\mp ~=~ - \hbar^2  \frac{d^2}{dx^2} +   V_\mp(x)~,
\end{equation}
with corresponding potentials $ V_\mp(x)=W^2\mp \hbar \;dW/dx$. 
These Hamiltonians  $H_{\mp}$ are intertwined:
\begin{equation}
A^-H_- = H_+A^- \qquad \mbox{and} \qquad  A^+H_+ = H_-A^+~.\label{eq:intertwining}
\end{equation}

Since these Hamiltonians are positive definite, the eigenenergies of $H_+$ and $H_-$ are non-negative. If either of the Hamiltonians has a zero-energy ground state, we say that we have unbroken supersymmetry. If neither Hamiltonian has a zero-energy groundstate, then SUSY is said to be broken. A given superpotential $W$ may have both broken- and unbroken- SUSY phases depending on the values of relevant parameters.

 For unbroken SUSY, without loss of generality, we choose the Hamiltonian $H_- = {A}^+ {A}^-$ to have a zero energy ground state; i.e., $E_{0}^{(-)}=0$. Then  Eq.~(\ref{eq:intertwining}) leads to the isospectral relations
\begin{equation}
	\frac{~~~{A}^- }{\sqrt{E^{+}_{n} }} ~\psi^{(-)}_{n+1} 
	= ~\psi^{(+)}_{n} ~~; ~~ ~~
	\frac{~~~{A}^+}{\sqrt{E^{+}_{n} }}~\psi^{(+)}_{n} 
	= ~  \psi^{(-)}_{n+1}~, \label{eq.isospectrality1}
\end{equation}
and
\begin{equation}
	E_{n+1}^{(-)} = E_{n}^{(+)} , \quad ~ n=0,1,2,\cdots~, \label{eq.isospectrality2}
\end{equation}
where $E_{n}^{(\pm)}$ and $\psi^{(\pm)}_{n}$ are, respectively, the eigenvalues and eigenfunctions of $H_\pm$.

Since $H_-\psi_0^{(-)}(x)={A}^+ {A}^- \psi_0^{(-)}(x) =0$, we have ${A}^- \psi_0^{(-)}(x) =0$, and hence
\begin{equation} \label{eq.groundstate}
	\psi_0^{(-)}(x,a) = {\mathcal N}\, e^{-\frac1\hbar \int_{x_0}^x W(x,a) dx}~,
\end{equation}
where $ {\mathcal N}$ is the normalization constant and $x_0$ is an arbitrary but finite point on the real axis. 
Note that $H_+$ and $H_-$ cannot both have normalizable ground states with zero energy,\footnote{If  $H_+$ also had a zero energy ground state, the corresponding wave function would be $\psi_0^{(+)}(x,a) = {\mathcal N}\, e^{+\frac1\hbar \int_{x_0}^x W(x,a) dx}=1/\psi_0^{(-)}(x,a)~$, which would not be normalizable.} so the ground state energy of $H_+$ must be positive.

Since $\psi_0^{(-)}$ must be normalizable for unbroken SUSY, we must have $\int_{x_0}^{L,R} W(x,a)\, dx =\infty$  which requires that $W$ be negative as $x\to L$ (the left boundary of the domain) and be positive as $x\to R$ (the right boundary). If the signs of $W$ at the limits of the domain are reversed, then $H_+$ would be the Hamiltonian corresponding to the zero-energy ground state.

If the asymptotic values of $W$ at $x\to {L,R}$ have the same sign, then the zero-energy state is non-normalizable and the system is in a broken SUSY phase. In this case, the ground state energy is positive for both potentials. For broken SUSY,  Eq.~(\ref{eq:intertwining})  gives 
\begin{equation}
E_n^B\equiv E^{-}_{n} =E^{+}_{n}, \quad \mbox{where}~ n=0,1,2,\cdots~\label{brokenintertwining}.
\end{equation}

\begin{figure}[tbh]
	\centering
	\includegraphics[width=0.5\linewidth]{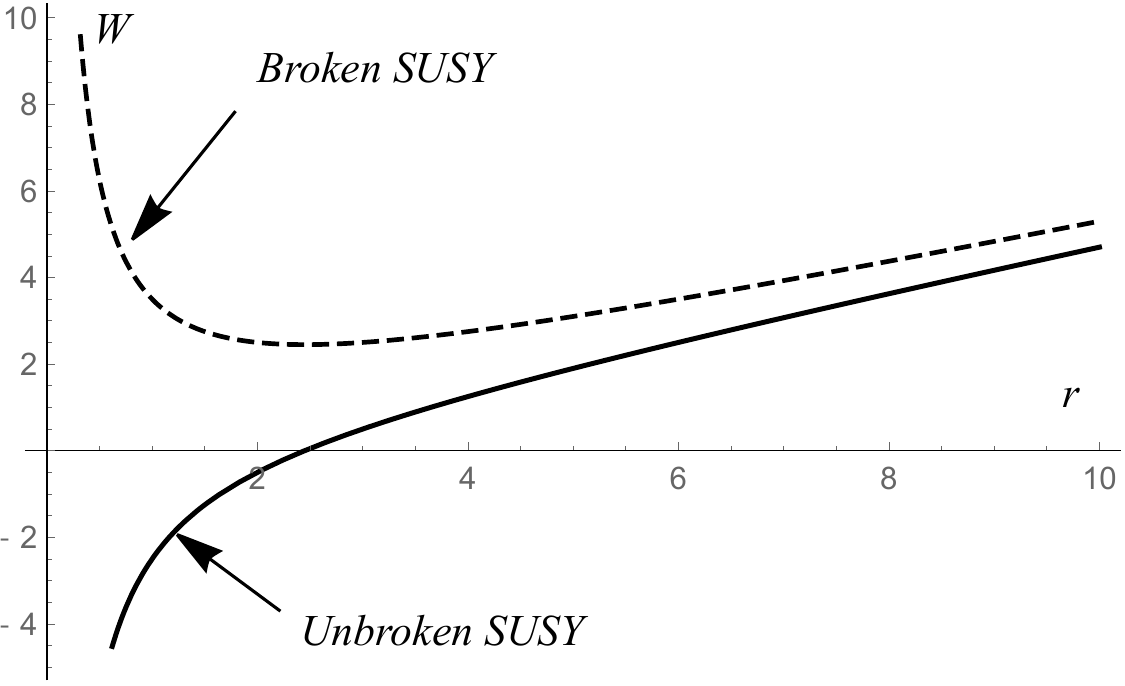}
	\caption{Broken ($\ell = -3$; dashed curve) and unbroken  ($\ell = 3$; solid curve) phases of the 3-D oscillator superpotential, in units such that $\omega = 1$. For the broken phase the superpotential $W$ is positive at both ends; in the unbroken phase, $W$ has opposite signs. Reprinted with permission from Gangopadhyaya, et al. \cite{Gangopadhyaya2021} \copyright 2021 IOP Publishing Ltd.}
	\label{fig:w-3d-o}
\end{figure}

\begin{figure}[tbh]
	\centering
	\includegraphics[width=\linewidth]{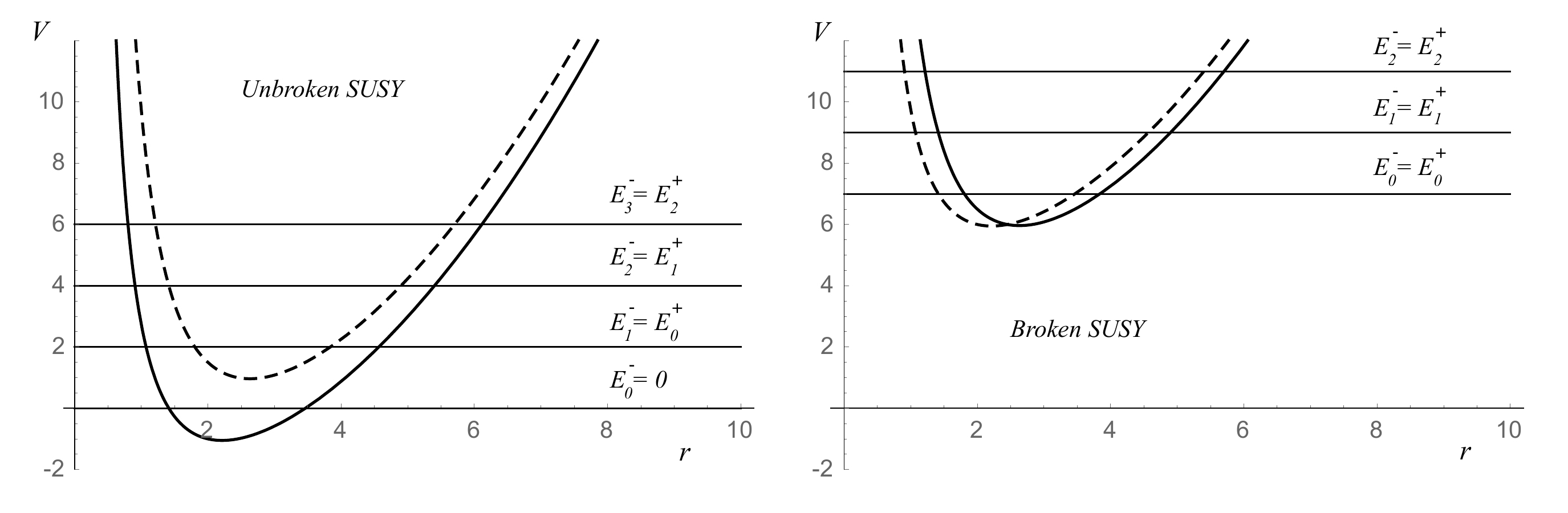}
	\caption{The potentials $V_+$ (dashed) and $V_-$ (solid) and their eigenenergies for the 3-D oscillator in the unbroken ($\ell = 3$; left plot) and broken ($\ell = -3$; right plot) phases, with units such that $\hbar = \omega = 1$. Reprinted with permission from Gangopadhyaya, et al.\cite{Gangopadhyaya2021} \copyright 2021 IOP Publishing Ltd.}
	\label{fig:vpmb-ub}
\end{figure}

Here we present an example of the 3-D harmonic oscillator (or radial oscillator) superpotential.
The superpotential is 
\begin{equation}
	\label{eq:W3DO}
	W(r,\omega,\ell) = \frac{\omega r}{2} -\frac{\ell}r~~;\qquad  0 < r < \infty~.
\end{equation}
 This superpotential exhibits both unbroken and broken phases, depending on the values of parameters $\omega$ and $\ell$.  
For $\omega >0$, the supersymmetry is unbroken for $\ell > 0$ and broken for $\ell < 0$. 
In Fig. \ref{fig:w-3d-o} we illustrate the two phases.
 In Fig.  \ref{fig:vpmb-ub}, we illustrate the radial oscillator partner potentials $V_{\pm}$ and their eigenvalues. In the unbroken phase the eigenvalues are the familiar $E_n^{-} = 2n\hbar\omega$, but in the broken phase they are \cite{Gangopadhyaya2001}  $E_n^B= (2n+1)\hbar\omega - 2\ell\omega$. 


\subsection{Shape Invariance}\label{sec.shapeinvariance}
In SUSYQM, a potential is called ``shape invariant'' if it satisfies the shape invariance condition  \cite{Infeld1951,Miller1968,Gendenshtein1983}:
\begin{eqnarray} 
	V_+(x, a_0) +g(a_0) =V_{-}(x, a_0) +g( a_1 ).
	\end{eqnarray} 
Here, $a_0$ and $a_1$ are parameters related by $a_1 = f(a_0)$. In terms of superpotential, this condition reads
\begin{equation} 
	W^2(x, a_0) + \hbar\, \frac{d\,W(x, a_0)}{dx} +g(a_0) = W^2(x, a_1 ) - \hbar \,
	\frac{d\,W(x, a_1 )}{dx}+g( a_1 ).\label{eq.SIC_W}
\end{equation} 
 In this paper we only consider additive, or translational shape invariance, where $a_1 = a_0+\hbar$.

For a system with unbroken supersymmetry, shape invariance guarantees exact solvability, as we will now show.
 Adding the kinetic energy terms on both sides of Eq. (\ref{eq.SIC_W}) yields
$$H_+(x,a_0)+g(a_0) = H_-(x,a_1)+g(a_1)~.$$
Thus, the ground states for the partner potentials are related by 
\[E_0^{(+)}(a_0)+g(a_0) = E_0^{(-)}(a_1)+g(a_1)~.\]
For unbroken SUSY, we automatically know the groundstate eigenvalue and eigenfunction for $H_-(x,a_1)$: 
\[
E_0^{(-)}(a_1) =0 \quad ;\quad \psi^{(-)}_{0}(x,a_1) = ~{\mathcal N}\, e^{-\frac1\hbar \int_{x_0}^x W(x,a_1) dx}~.
\]
Thus, using Eq. (\ref{eq.isospectrality2}) and (\ref{eq.groundstate}), we obtain
\begin{equation}\label{eq.firstexcitedstate}
	E_1^{(-)}(a_0) = g(a_1)- g(a_0) \quad ;\quad
	\psi^{(-)}_{1}(x,a_0) 
	=\frac{~~~{A}^-(a_0) }{\sqrt{E^{+}_{n}(a_0) }} 
	\quad {\mathcal N}\, e^{-\frac1\hbar \int_{x_0}^x W(x,a_1) dx}~.
\end{equation}
Iterating this procedure determines the eigenvalues and eigenfunctions for excited states of $H_-(x,a_0)$:
\begin{eqnarray}
	E_n^{(-)}(a_0)&=&g(a_n)-g(a_0)~, \label{eq:En} \\
	\nonumber\\
	\psi^{(-)}_{n}(x,a_0)&=&
	\frac{{\mathcal A}^+{(a_0)} {\mathcal A}^+{(a_1)}  \cdots  {\mathcal A}^+{(a_{n-1})}}
	{\sqrt{E_{n}^{(-)}(a_0)\,E_{n-1}^{(-)}(a_1)\cdots E_{1}^{(-)}(a_{n-1})}}
	\quad \psi^{(-)}_0(x,a_n)\\
	\nonumber\\
	&=&
	\frac{{\mathcal A}^+{(a_0)}  {\mathcal A}^+{(a_1)}  \cdots  {\mathcal A}^+{(a_{n-1})}}
	{\sqrt{E_{n}^{(-)}(a_0)\,E_{n-1}^{(-)}(a_1)\cdots E_{1}^{(-)}(a_{n-1})}}
	\quad {\mathcal N}\, e^{-\frac1\hbar \int_{x_0}^x W(x,a_n) dx}
	~.
\end{eqnarray}
Thus, unbroken SUSY provides the ground state energy and wavefunction, from which shape invariance allows us to derive higher energy levels. Therefore, all shape-invariant potentials are exactly solvable in their unbroken phase.

In the next section, we show that shape invariance constrains the possible functional forms for  $W$; we categorize SIPs based on their form and examine connections between these categories.

\section{Categorizing Shape-Invariant Potentials\label{sec:SIPs}}
	
The shape invariance condition, Eq. ({\ref{eq.SIC_W}}), is a difference-differential equation. It connects superpotentials $W(x,a)$ and $W(x,a+\hbar)$, and their derivatives, for two different values of the parameter $a$. 	By requiring that this relation holds for all values of $\hbar$, including the limit $\hbar \rightarrow 0$, we expand Eq. ({\ref{eq.SIC_W}}) in powers of $\hbar$. By equating powers of $\hbar$, we express shape-invariance through a set of differential equations. These differential equations allow us to categorize SIPs.

Until recently, the known shape-invariant superpotentials\cite{Infeld1951, Dutt1988} did not depend explicitly on  $\hbar$; we now call these superpotentials ``conventional.''
In 2008, researchers discovered \cite{Quesne2008,Quesne2009} and generalized \cite{Odake2009, Odake2010, Tanaka2010, Odake2011, Odake2013} a new family of ``extended'' superpotentials, which have been shown\cite{Bougie2010,Bougie2012} to depend explicitly on $\hbar$.

\subsection{Conventional SIPs and Their Classes\label{sec:conventional}}

We now expand Eq. ({\ref{eq.SIC_W}}) in powers of $\hbar$. Assuming the superpotential is conventional, $\hbar$ can  only appear in $W$ through evaluation at $a+\hbar$.  At the lowest order, \cite{Gangopadhyaya2008} we get
\begin{equation}
	W \, \frac{\partial W}{\partial a} - \frac{\partial W}{\partial x} + \frac12 \, \frac{d g}{d a}~=0, \label{eq.PDE1}
\end{equation}
and all higher order equations reduce to \cite{Bougie2010,Bougie2012}
\begin{equation}
	\frac{\partial^{3}}{\partial a^{2}\partial x} ~W(x,a) = 0~.
	\label{eq.PDE2}
\end{equation}
Eq. (\ref{eq.PDE2}) implies that conventional superpotentials $W(x,a)$ must be of the form:
\begin{equation}
	W(x,a) = a f_1(x) + f_2(x)+u(a) ~. \label{eq.PDE2solution}
\end{equation}

This structure, originally conjectured by Infeld et al. \cite{Infeld1951} in piecemeal  form in the 1950s, has since served as an ansatz for other researchers \cite{Carinena2000a, Carinena2000b, Cheng2003}. We divide conventional potentials into three classes as follows:
\begin{itemize}
	\item Class I: $f_1(x)$ is a constant, so the only $x$-dependence term in $W$ is $f_2(x)$.
	\item Class II: $f_2(x)$ is a constant, so the only $x$-dependence term in $W$ is $a f_1(x)$.
	\item Class III: Neither $f_1(x)$ nor $f_2(x)$ are constant, so that both $a f_1(x)$ and $f_2(x)$ contain functional $x$-dependence.
\end{itemize} 

The authors of Ref.~\cite{Bougie2010} proved that the known list of conventional potentials are the solutions of Eqs. (\ref{eq.PDE1}) and (\ref{eq.PDE2}), and that the list is complete.  We include this list in Appendix \ref{AppendixA}, together with the class to which each potential belongs.

\subsection{Extended SIPs\label{sec:extended}}

For extended superpotentials, we expand $W$  in powers of $\hbar$, as
\begin{equation}
	W(x, a, \hbar) = \sum_{i=0}^\infty \hbar^i W_i(x,a)~;
	\label{W-hbar}
\end{equation}
this can be written as
\begin{equation}
	W(x,a,\hbar)=W_0(x,a)+W_h(x,a,\hbar),
\end{equation}
where the ``kernel'' $W_0$ does not depend on $\hbar$ and $W_h=\sum_{i=1}^\infty \hbar^i W_i(x,a)$ includes all explicit $\hbar$-dependence.

Substituting  Eq. (\ref{W-hbar}) into Eq. (\ref{eq.SIC_W}), and requiring that Eq. (\ref{eq.SIC_W}) hold for any value of $\hbar$, the coefficients of the series must separately equal zero for each power of $\hbar$. 
To first order in $\hbar$, this gives:
\begin{eqnarray}
	W_0 \, \frac{\partial W_0}{\partial a} - \frac{\partial W_0}{\partial x} + \frac12 \, \frac{d g}{d a}~=0,
	\label{lowestorder}
\end{eqnarray}
and for order of $\hbar^j$ where $j \geq 2$, we get
\begin{equation}
	2\,\frac{\partial W_{j-1}}{\partial x}
	-\sum_{s=1}^{j-1} \sum_{k=0}^s \frac{1}{(j-s)!}
	\frac{\partial^{j-s}}{\partial a^{j-s}} W_k\, W_{s-k} 
	+  
	\sum_{k=2}^{j-1} \frac{1}{(k-1)!}
	\frac{\partial^{k}\, W_{j-k}}{\partial a^{k-1} \,\partial x}=0 
	~.
	\label{higherorders}
\end{equation}
Since $g(a)$ appears in  Eq. (\ref{lowestorder}) but not in Eq. (\ref{higherorders}), we see\cite{Bougie2010,Bougie2012} that 
the energy spectrum is determined solely by $W_0$.
	
All conventional SIPs satisfy Eq.~(\ref{lowestorder}), so each conventional SIP could serve as a kernel for extended SIPs. We note that conventional potentials have  $W_i=0$ for all $i>0$; in this case, Eq. (\ref{higherorders}) reduces to  Eq. (\ref{eq.PDE2}), which is solved by Eq.~(\ref{eq.PDE2solution}).  
It is  possible, in principle, for $W_0$ to be a function that is not a conventional potential, as long as it satisfies Eq.~(\ref{lowestorder}) and all $W_i$ satisfy  Eq. (\ref{higherorders}).  However, every known extended SIP has a conventional SIP as its kernel, and thus each known extended SIP is isospectral with its conventional kernel.

As an example, the authors  of Refs. \cite{Bougie2010,Bougie2012} began with the harmonic oscillator superpotential $W_0(r,\ell) = \frac{1}{2}\, \omega r-\frac{\ell}{r}$ as a kernel, and determined $W_i(r,\ell), \,{\rm for \, all} ~i > 0$ from Eq. (\ref{higherorders}), to generate the extended superpotential 
\begin{equation}
		\label{eq:Quesne-ext}
		W(r, \ell, \hbar)= \frac{1}{2}\, \omega r-\frac{\ell}{r} +\left(\frac{2 \omega r \hbar}{\omega r^{2}+2 \ell-\hbar}-\frac{2 \omega r \hbar}{\omega r^{2}+2 \ell +\hbar}\right)~.
\end{equation}
This reproduces the extended superpotential first discovered by Quesne in Ref.~\cite{Quesne2008}, with $\hbar=1$.
In the next section, we will discuss the connections between all known additive SIPs.

\subsection{Interconnection of all additive shape invariant potentials}
\begin{figure}[htpb]
	\centering
	\includegraphics[width=0.9\linewidth]{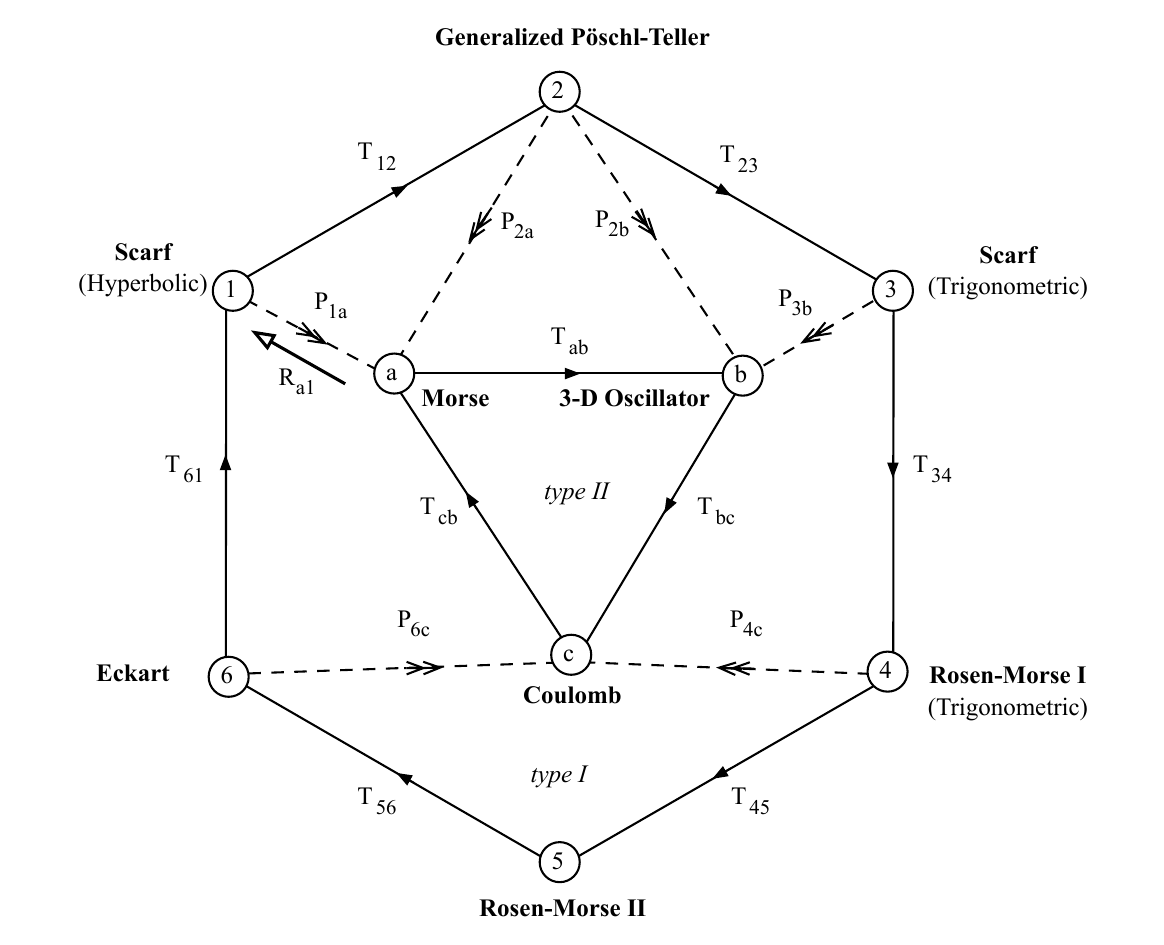}
	\caption{All conventional SIPs are interconnected. PCTs are represented by plain lines with arrows, projections by dashed lines with double arrows, and the restricted extension $R_{a1}$ by a bold arrow. Reprinted with permission from Mallow, et al. \cite{Mallow2020} \copyright 2019 Elsevier B.V.
	\label{fig:Hexagon}}
\end{figure}

Appendix \ref{AppendixA} lists all conventional SIPs.  
The Schr\"odinger equation for six of these potentials reduces to hypergeometric differential equations, which we call type-I; these occupy the outer hexagonal ring\cite{Mallow2020} in Fig. \ref{fig:Hexagon}.  The remaining three \footnote{We consider the one-dimensional harmonic oscillator to be a simplified case of the 3-D oscillator where \(\ell=0\), thus removing the singularity at the origin and  changing the domain to the entire real axis.} potentials lead to confluent hypergeometric differential equations. We call them type-II and they are situated at the corners of the triangle in the diagram. These potentials are connected by point canonical transformations ($T_{ij}$), projections ($P_{\alpha\beta}$), and a restricted extension  ($R_{a1}$) as described below.
\subsubsection{Point Canonical Transformations (PCT)}
A PCT transforms a given Schr\"odinger equation to a new one by a change of the independent variable $x$ and a corresponding multiplicative transformation of the wavefunction \cite{Bhattacharjie1962,De1992}. 
A change of variable from $x \to z$, where $x=u(z)$, must be accompanied by a change in wave function  $\psi \to \tilde{\psi}$, where $\psi(x) = \tilde{\psi}(z)\,\sqrt{du/dz} $.
This transforms the Schr\"odinger equation 
\begin{equation}
	-\frac{d^2 \psi(x)}{dx^2} + V(x,a_i)\, \psi(x)  = 
	E(a_i) \psi(x)~,
	\label{eq.1}
\end{equation}
into
\begin{equation}
	\left[ 	-\frac{d^2 }{dz^2}
	+ \left\{ 
	{u}'^{\,2}\, \left[ V(u(z),a_i) - E(a_i) \right] + 
	{\frac 12}
	\left( \frac{3{u''}^{\,2}}{ 2 {u'}^{\,2}} - \frac{{u'''}}{{u'}}  \right)
	\right\} 
	\right] \tilde{\psi} \left( z \right) = 0,
	\label{eq:second_order}
\end{equation}\\
where primes denote derivatives with respect to $z$, so that ${u}' \equiv du/dz$. For Eq. (\ref{eq:second_order}) to be a Schr\"odinger equation, the expression 
${u'}^{\,2} \, \left[ V(u(z),a_i) - E(a_i)\right]$ must have a term that is independent of $z$, corresponding to the energy, thus constraining possible choices for the function $u(z)$.  

As Fig. \ref{fig:Hexagon} shows, the following PCTs map each conventional SIP to another SIP  of the same type \cite{Cooper1989, De1992, Levai1994}:
$
T_{12}\!:\! \{x\to r+ i \pi /2 \},\, 
T_{23}\!:\! \{ r\to i x+ i \pi /2\},\, 
T_{34}\!:\! \{x\to \cos ^{-1}({\rm cosec } x) \},\, 
T_{45}\!:\! \{ x\to \pi /2 +i x \}, \,
T_{56}\!:\! \{ x\to -r+ i \pi/2\},\, 
T_{61}\!:\! \{ r\to \coth ^{-1}(i \sinh x) \},\, 
T_{ab}\!:\! \{x\to -2 \ln r\},\, 
T_{bc}\!:\! \{ r\to \sqrt z\},\, 
$
and
$
T_{ca}\!:\! \{ r\to \exp (-x)\}\,.
$

\subsubsection{Projections}

Although they provide connections within a type, PCTs do not connect type-I SIPs to type-II. However, ``projections'' transform type-I to type-II. Type-I superpotentials correspond to the hypergeometric differential equation, which has three regular singular points. The equation reduces to a confluent hypergeometric equation corresponding to type-II in the appropriate limit in which two of the singularities merge. 
Table~3
in Appendix \ref{AppendixB}, shows the appropriate limits corresponding to these projections, which are illustrated by dashed lines in Figure \ref{fig:Hexagon}. 

\subsubsection{Isospectral Extension}
Point canonical transformations connect SIPs within a type (type-I or type-II), but cannot go from one type to another. Projections reduce one parameter, so they only go from type-I to type-II. As we have already seen in Sec. \ref{sec:extended}, adding $\hbar$-dependent terms satisfying Eq. (\ref{higherorders}) to a conventional SIP can produce an extended SIP. 

Additionally, the differential equations required by the $\hbar$-expansion can generate a ``restricted'' isospectral extension that produces a second parameter in a type-II SIP, creating a pathway from type-II to type-I. This was done in Ref.~\cite{Mallow2020} to generate a path from Morse to Scarf.

We begin with a Morse superpotential as kernel: $W_0=-a - e^{-x}$. Setting $W_{2k-1}=0$ for all positive integers $k$, we find that
\begin{equation*}
	W_{2k}=
	\left(- Q\right)^{k-1} e^{-(2k-1) x} \left( 2 P + Q e^{-2 x}
	+ 2 a ~  Q e^{- x}\right)~
\end{equation*}
satisfies Eq. (\ref{higherorders}) for all positive integers $k$,  where $P$ and $Q$ are constant parameters. 
The infinite sum $\sum_{j=0}^\infty \hbar^j W_j(x,a)$, produces a shape-invariant superpotential that is an isospectral extension of Morse:
\begin{equation}
	W(x,a,\hbar) = -a -e^{- x}  +
	\frac{\hbar^2 \left( 2 Pe^{x} + 2 a\, Q + Q\, e^{- x}
		\right)}{e^{2 x}+Q\,\hbar^2}~.
	\label{newpotential}
\end{equation}
Substituting $W(x,a,\hbar)$ into Eq. (\ref{eq.SIC_W}) verifies that this procedure maintains shape-invariance, as it must, where $g(a) =
-a^2$. The corresponding energies are $E^{(-)}_n =
g(a+n\hbar)-g(a) = a^2 - (a+n\hbar)^2$. 

This particular extension has unusual features. The $\hbar$-dependence in $W(x,a,\hbar)$ is superficial and can be absorbed by redefinition of parameters as follows:
\begin{equation}
	\label{toScarf}
	\hbar^2 P \to P^{\,\prime} ,~
	\hbar^2 Q \to e^{2\beta}, ~
	(2 P^{\,\prime} -1)\,e^{-\beta} \to 2 B,~
	-a \to A,~
	x-\beta \to x ~.
\end{equation}
These transformations convert $W(x,a,\hbar)$ into the Scarf hyperbolic, a conventional type-I superpotential:
\begin{equation}
	-a - e^{- x}  +
	\frac{\hbar^2 \left( 2 P\,e^{x} + 2a\, Q + Q\, e^{- x}
		\right)}{e^{2 x}+Q\,\hbar^2}~
	\to W_{\rm Scar\!f} = A \tanh x + B\,\textrm{sech}\, x~.
\end{equation}
Thus, we have constructed a ``restricted extension'' $R_{a1}$ which provides a path from a type-II to a type-I superpotential. 
Thus all SIPs are connected by reversible paths.

\section{SWKB Exactness for Conventional Potentials in the Unbroken Phase}
\label{sec:UBSWKB}

In Ref.~\cite{Gangopadhyaya2020} , the authors proved that exactness of the SWKB quantization condition, Eq. (\ref{eq:swkb}), for conventional SIPs follows from additive shape invariance. In this section, we revisit the proofs of Ref.~\cite{Gangopadhyaya2020} using a more direct path. 

Henceforth, we denote differentiation of any function $f(a,x)$ with respect to $a$ by a dot, $\dot{f}=\frac{\partial f}{\partial a}$, and with respect to $x$ by a prime,  ${f'}=\frac{\partial f}{\partial x}$. We will also use $E_n$ (without superscript)  for $E_n^-$.

For a given superpotential $W$ and energy $E_n$ the SWKB quantization condition, Eqs. (\ref{eq:swkb}) and (\ref{eq:bswkb}),  is exact if $I_S = n \pi\hbar$  for unbroken SUSY and $I_S = (n+1/2) \pi\hbar$ for broken SUSY.
We will prove that SWKB exactness directly follows from the following minimal set of assumptions for a conventional superpotential $W$:
\begin{itemize}
\item From Eq.~(\ref{eq.PDE1}), we require that $W\dot{W}-W'+\dot{g}/2=0$.
\item From  Eq.~(\ref{eq.PDE2}), we require that $\ddot{W}'=0$.
 The general solution is \\ $W(x,a) = a f_1(x) + f_2(x)+u(a)$.
\item To avoid level crossing, we require that $E_{n+1}>E_n$ for all bound states.
\end{itemize}

We distinguish between the broken and unbroken phases of the superpotential based on the behavior at the limits of the domain.
\begin{itemize}
	\item For the unbroken phase, $W$ has opposite signs at the two ends of the domain. From Eq.~(\ref{eq:En}), $E_n=g(a_n)-g(a_0)$, where $a_n=a_0+n\hbar$, such that $E_0=0$. To avoid level crossing $\dot{g}(a)|_{a_n}>0$ for all bound states. In this section we evaluate \(I_S\) and show that it is exact. 	
	\item For the broken phase, $W$ has the same sign at the two ends of the domain. Therefore, $E_n-W^2$ can only have two roots if $W$ has at least one extremum within the domain.
	In the broken phase the groundstate energy is not known {\it a priori}, therefore, in the next section we find the energies using a discrete SI symmetry and prove the exactness of \(I_S\).
	\end{itemize}

We begin with the general form $W(x,a) = a f_1(x) + f_2(x)+u(a)$ and identify three classes: 
\begin{enumerate}
	\item[] \textbf{Class I}: $f_1$ is a constant;
	\item[] \textbf{Class II}: $f_2$ is a constant;
	\item[] \textbf{Class III}: Neither $f_1$ nor $f_2$ are constants.
\end{enumerate}
Note that \(f_1\) and \(f_2\) cannot both be constant because that would result into a trivial superpotential. 
We now examine each of these classes, and show that SWKB is exact.

\subsection{Class I Potentials: $f_1$ is a constant}\label{Sec:classISWKB}
 Because $f_1$ is a constant, we absorb it into $u(a)$, so $W(x,a)=f_2(x)+u(a)$. From Eq.~(\ref{eq.PDE1}) $\big[f_2(x)+u(a)\big] \dot{u}(a) - f_2'(x)  + \frac12 \dot{g}(a)~=0$. Since $f_2$ is not constant, we must have 
  \begin{equation}
 	-\left(f_2\dot{u}-f_2' \right)=u\dot{u}+\dot{g}/2=\omega/2 
 \end{equation}
 for some constant $\omega$.  
 Then \(\dot{u} = \alpha\) so \(u=\alpha a + \beta\) for some constants \(\alpha\) and \(\beta\). We absorb \(\beta\) in \(f_2\), giving 
  \begin{equation}
 	W=f_2 +\alpha a,
 \end{equation}
 \begin{equation}
 	\label{eq:classI}
 	f_2'=\alpha f_2 +\omega/2, 
 \end{equation}
 \begin{equation}
	\dot{g}=\omega-2\alpha^2a.
\end{equation}
  We now consider two subclasses: Subclass IA, where $\alpha=0$, and Subclass IB, where $\alpha\neq0$.
 
\subsubsection{Subclass IA: $W(x,a)=f_2(x)$}

In this case, $W=f_2$, and Eq.~(\ref{eq:classI}) mandates that $W'=f_2'=\omega/2$. 
Therefore, we note the following:
\begin{itemize}
	\item \textbf{Unbroken SUSY:} Since $W=f_2$, which is monotonic in $x$ and whose range is $-\infty<f_2<\infty$, this superpotential is always in the unbroken phase. In this case, $E_n=\omega n \hbar$, where $\omega>0$.
	\item \textbf{Broken SUSY:} Due to the above argument, there is no broken phase.
\end{itemize}
We therefore have
\begin{eqnarray}
	I_{S}&=& \int_{x_L}^{x_R} \sqrt{E_n-W^2(x,a)}\quad {\rm d}x = \int_{-\sqrt{E_n}}^{\sqrt{E_n}}\frac{ \sqrt{E_n-W^2}\quad {\rm d}W}{W'}\nonumber\\
	&=&\frac{2}{\omega}\int_{-\sqrt{E_n}}^{\sqrt{E_n}} \sqrt{E_n-W^2}\quad {\rm d}W.
\end{eqnarray}
The integral is of the form of $I_{0}$ from  Appendix \ref{AppendixC}, with $y_1=-\sqrt{E_n}$ and $y_2=\sqrt{E_n}$. Therefore,  
\begin{eqnarray}
	I_{S}&=&\frac{2}{\omega} \frac{\pi}{8}\left(\sqrt{E_n}+\sqrt{E_n}\right)^2=\frac{2}{\omega} \frac{\pi}{8}\left(2\sqrt{\omega n\hbar}\right)^2=n\pi\hbar.
\end{eqnarray}

\subsubsection{Subclass IB: $W(x,a)=f_2(x)+\alpha a$, $\alpha\neq0$}
In this case, we have the form $W=f_2(x)+\alpha a$, where $\alpha\neq0$. We define
\begin{equation}
	y\equiv f_2+\omega/(2\alpha) = W+\dot{g}/(2\alpha),
\end{equation}
which implies $y'=f_2'=W'=\alpha y$. For nontrivial potentials, $y$ and $y'$ cannot cross zero within the domain and thus both must have fixed signs.

Therefore, we note the following regarding the broken and unbroken phases:
\begin{itemize}
	\item \textbf{Unbroken SUSY:} We note that $W=y-\dot{g}/(2\alpha)$, where $\dot{g}=\omega-2\alpha^2a>0$, and $y$ cannot cross zero within the domain. Therefore, SUSY is unbroken when $y$ and $\alpha$ have the same sign; {\it i.e.,} $\alpha y >0$.  The energies are given by
		$E_n=\omega n \hbar - \alpha^2\left(n^2\hbar^2 + 2 a_0 n \hbar \right).$
	\item \textbf{Broken SUSY:}
	Since $W$ is monotonic in $x$, the BWSKB integral does not have two integration limits.
\end{itemize}
For the unbroken phase, the integration limits occur when $W=\pm\sqrt{E_n}$, which corresponds to
$y_L=-\sqrt{E_n}+\frac{\dot{g}}{2\alpha}$ and $y_R=\sqrt{E_n}+\frac{\dot{g}}{2\alpha}.$ Since $y$ and $\alpha$ must have the same sign, this requires $y'>0$.

The SWKB quantization condition thus becomes  
\begin{equation*}
	I_{S} = \int_{x_1}^{x_2} \sqrt{E_n-W^2(x,a)}\quad {\rm d}x = \int_{y_L}^{y_R}\frac{ \sqrt{E_n-\left[y-(\frac{\dot{g}}{2\alpha})^2\right]}\quad {\rm d}y}{\alpha y}.
\end{equation*}
which simplifies to 
\begin{equation*}
	I_{S}=\frac{1}{\alpha} \int_{y_L}^{y_R}\frac{ \sqrt{(y_R-y)(y-y_L)}\quad {\rm d}y}{ y}.
\end{equation*}
For $y>0$ the integral is of the form $I_{1a}$ while for $y<0$ it is of the form $I_{1b}$ from  Appendix \ref{AppendixC}.
We get
\begin{eqnarray}
	I_{S}&=&\frac{1}{\alpha} \left[\frac{\pi}{2}(y_L+y_R)-\varepsilon\pi\sqrt{y_L\, y_R}\right],\label{eq:meh}
\end{eqnarray}
where $\varepsilon=1$ if $y>0$ and $\varepsilon=-1$ if $y<0$.
Substituting the expressions for $y_L$ and $y_R$, along with the known values of $E_n$ and $\dot{g}$, Eq. (\ref{eq:meh}) becomes
\begin{equation}
	I_{S}=\frac{1}{\alpha}\left[\frac{\pi}{2\alpha} \left({\omega-2\alpha^2 a}\right)-\frac{\pi}{2\alpha}\left(\omega-2\alpha^2a-2\alpha^2 n\hbar)\right)\right] =\pi n \hbar.
\end{equation}

\subsection{Class II Potentials: $f_2$ is a constant}\label{Sec:classIISWKB}
For Class II superpotentials we absorb the constant $f_2$ into $u(a)$ and write the superpotential as $W(x,a)=a f_1(x)+u(a)$. Differentiating Eq.~(\ref{eq.PDE1}) twice with respect to $a$ yields $	f_1\left(3\ddot{u}+a\dddot{u} \right)+ 3\dot{u}\ddot{u}+u\dddot{u}+\dddot{g}/2=0.$
Since $f_1$ cannot be constant and $u(a)$ is independent of $x$, this requires that $3\ddot{u}+ a \dddot{u}=0$, so \(u = B/a + \mu a + \nu\). We absorb \(\mu a\) in  \(a f_1\) so $u(a)=B/a+\nu$ for constants $\nu$ and $B$.

Inserting the form 
$W=f_1 + \nu+ B/a$ into Eq.~(\ref{eq.PDE1}) yields $$	a \left( f_1^2 -f_1'\right) + \nu f_1 - \frac{B}{a^3}-\frac{B\nu}{a^2}+\frac{\dot{g}}{2}=0.$$
Since $f_1$ is not constant, $\nu=0$. Thus, $
	a\left(f_1^2 - f_1'\right) = \frac{B^2}{a^3}-\frac{\dot{g}}{2} = a \lambda$
for some constant $\lambda$.
This leads to 
\begin{equation}
	W=a f_1 + B/a,
\end{equation}
\begin{equation}
	f_1'=f_1^2-\lambda, 
\end{equation}
\begin{equation}
	\dot{g}=\frac{2B^2}{a^3}-2 a \lambda.
\end{equation}
Consequently, Class II splits into Subclass IIA, where $\lambda=0$, and Subclass IIB where $\lambda\neq0.$

\subsubsection{Subclass IIA: $W(x,a)=a f_1(x)+B/a$; $f_1'=f_1^2$; $\lambda=0$}

 In this case,  $f_1$ and $f_1'$ cannot cross zero for nontrivial potentials. We therefore note the following regarding the unbroken and broken phases:
\begin{itemize}
	\item \textbf{Unbroken SUSY:} Since $\dot{g}=2B^2/a^3>0$, to avoid level crossing we must have $a>0$, and the energy levels are given by $E_n=B^2\left[\frac{1}{a^2}-\frac{1}{(a+n\hbar)^2}\right]$. Additionally, $B$ and $f_1$ must have opposite signs. 
	\item \textbf{Broken SUSY:}
Since $W$ is monotonic in $x$, the BWSKB integral does not have two integration limits.
\end{itemize}

For the unbroken phase we change integration variables in \(I_S\) so that $y= a f_1$, and $y'=y^2/a$. The SWKB condition therefore becomes
 \begin{equation}
	I_{S}= \int_{x_L}^{x_R} \sqrt{E_n-W^2(x,a)}\, {\rm d}x =
	 a\int_{y_L}^{y_R}\frac{\sqrt{E_n-(y+\frac{B}{a})^2}}{y^2}\,{\rm d}y~,
\end{equation}
where the integration limits are $y_L=-\sqrt{E_n}-B/a$ and $y_R=\sqrt{E_n}-B/a$.
This simplifies to 
 \begin{eqnarray}
	I_{S}&=& a \int_{y_L}^{y_R}\frac{\sqrt{(y_L-y)(y-y_R)}}{y^2}\,{\rm d}y.
\end{eqnarray}
Since $y$ cannot change sign and $a>0$, we have one of two cases. Either $y_R>y_L>0$, which requires $B<-a\sqrt{E_n}<0$, or  $y_L<y_R<0$, which requires $B>a\sqrt{E_n}>0$.
For $y>0$ the integral is of the form $I_{2a}$ while for $y<0$ it is of the form $I_{2b}$ from  Appendix \ref{AppendixC}.
We get
\begin{equation}
	\label{eq:meh1}
	I_S=\frac{a \pi}{2}\left(\frac{\varepsilon y_L+\varepsilon y_R-2\sqrt{y_L y_R}}{\sqrt{y_1 y_2}}\right),
\end{equation}
 where $\varepsilon=1$ when $y>0$ and $\varepsilon=-1$ when $y<0$. 
Substituting the integration limits and the energy levels, Eq. (\ref{eq:meh1}) yields 
 \begin{equation}
	I_{S} = \pi a(a+n\hbar)\left(\frac{1}{a}-\frac{1}{a+n\hbar}\right)
	= n \pi \hbar.
\end{equation}

\subsubsection{Subclass IIB: $W(x,a)=a f_1(x)+B/a$; $f_1'=f_1^2-\lambda$; $\lambda\neq0$}

 In this case,  $f_1^2\neq\lambda$ and $f_1'\neq0$ within the domain for nontrivial potentials. We note the following regarding the unbroken and broken phases:

\begin{itemize}
	\item \textbf{Unbroken SUSY:} In this case,	$\dot{g}=\frac{2B^2}{a^3}-2 a \lambda>0$, and $E_n=\frac{B^2}{a^2}- \frac{B^2}{(a+n\hbar)^2}+\lambda\left[a^2-(a+n\hbar)^2\right].$ 
	\item \textbf{Broken SUSY:}
Since $W$ is monotonic in $x$, the BWSKB integral does not have two integration limits.
\end{itemize}
For the unbroken phase, we define \(y \equiv f_1/\sqrt{\varepsilon_1 \lambda}\) and obtain
$
	y' = \sqrt{\varepsilon_1 \lambda}\left(y^2-\varepsilon_1\right)
$,
where $\varepsilon_1=1$ when $\lambda>0$ and $\varepsilon_1=-1$ when $\lambda<0$; thus, $\varepsilon_1 \lambda=|\lambda|$.
Therefore
\begin{equation}
W=a y\sqrt{\varepsilon_1 \lambda} +B/a~; \quad W' = a \varepsilon_1 \lambda
\left(y^2-\varepsilon_1\right)~.
\end{equation}
Note that $y^2$ cannot cross $\varepsilon_1$ within the domain for non-trivial potentials; $W'$ and $y'$ cannot cross zero and the sign of \(W'\) is the same as that of \(ay'\). 

We define $\varepsilon_2$ such that $\varepsilon_2=1$ when $a>0$ and $\varepsilon_2=-1$ when $a<0$, and $\varepsilon_3$ such that $\varepsilon_3=1$ when $y'>0$ and $\varepsilon_3=-1$ when $y'<0$.  
We change our integration variable of the SWKB quantization condition such that 
 \begin{equation*}
	I_{S}=\int_{x_L}^{x_R} \sqrt{E_n-W^2(x,a)}\,{\rm d}x = 
 \frac{\varepsilon_3}{\sqrt{\varepsilon_1\lambda}}\int_{y_L}^{y_R}\frac{\sqrt{E_n-(a y\sqrt{\varepsilon_1 \lambda} +B/a)^2}}{y^2-\varepsilon_1}\,dy~,
\end{equation*}
where
\begin{equation*}
 y_L=\frac{1}{a\sqrt{\varepsilon_1 \lambda}}\left(-\varepsilon_2\sqrt{E_n}-\frac{B}{a}\right)~; \quad
 y_R=\frac{1}{a\sqrt{\varepsilon_1\lambda}}\left(\varepsilon_2\sqrt{E_n}-\frac{B}{a}\right)~.
\end{equation*}
This simplifies to
 \begin{eqnarray}
	I_{S}&=&  a \varepsilon_2 \varepsilon_3\int_{y_L}^{y_R}\frac{\sqrt{\left(y_R-y \right)\left( y-y_L \right)}}{y^2-\varepsilon_1}\, dy~.
	\label{eq.class2b}\end{eqnarray}
	
We evaluate the integral in Eq.(\ref{eq.class2b}) based on the signs of $\lambda$ and $a$.

\medskip
\noindent{\bf Case IIB1: $\lambda<0$.}
	
	For this case, because \(\dot{g}>0\) we have $\frac{B^2}{a^3}> a \lambda$; hence $a>0$, so $\varepsilon_1=-1$ and $\varepsilon_2=1$. Also, $y'=\sqrt{- \lambda}\left(y^2+1\right)>0$, so $\varepsilon_3=1.$
With these values, Eq.(\ref{eq.class2b}) becomes
 \begin{eqnarray}
	I_{S}&=&  a \int_{y_L}^{y_R}\frac{\sqrt{\left(y_R-y \right)\left( y-y_L \right)}}{y^2+1}\,dy = a I_3~,
	\label{eq.class2b-1}\end{eqnarray}
where $I_3$ is given in Appendix \ref{AppendixC}. This evaluates to
	 \begin{eqnarray}
		I_{S}&=&  \frac{a \pi}{\sqrt{2}}\left[\sqrt{(1+y_L^2)(1+y_R^2)}-y_L y\,_R +1\right]^{1/2} -\pi a \nonumber\\
		&=& \frac{a \pi}{\sqrt{2}}\left[\sqrt{\frac{\left[B^2-\lambda(a+n \hbar)^4\right]^2}{a^4(a+n \hbar)^4\lambda^2}}-\frac{-B^2-n\hbar(a+n\hbar)^2(2a+n\hbar)\lambda}{a^2(a+n\hbar)^2\lambda}\right]^{1/2}-\pi a\nonumber\\
		&=& \pi \sqrt{(a+n\hbar)^2} - \pi a
		= n\pi\hbar~.
		\label{eq.class2b0}\end{eqnarray}

\medskip
\noindent{\bf Case IIB2: $\lambda>0$ and $a<0$. }
	
		For this case, $\varepsilon_1=1$ and $\varepsilon_2=-1$; moreover $\frac{B^2}{a^3}> a \lambda$ requires $0<B^2<a^4\lambda$. We note that $(y_L + \, y_R)^2=\frac{4B^2}{a^4\lambda}<4. $ Since $y^2$ cannot cross 1 within the domain, this means that $-1<y_L<y_R<1$, so $y'<0$ and $\varepsilon_3=-1.$ 
		Therefore, 
		 \begin{eqnarray}
			I_{S}&=&-  a \int_{y_L}^{y_R}\frac{\sqrt{\left(y_R-y \right)\left( y-y_L \right)}}{1-y^2}\,dy =- a I_4~,
		\end{eqnarray}
	where $I_4$ is given in Appendix \ref{AppendixC}. We obtain
		 \begin{eqnarray}
		I_{S}
		&=& \frac{a \pi}{2}\left[\sqrt{(1-y_L)(1-y_R)}+\sqrt{(1+y_L)(1+y_R)}-2\right]\nonumber\\
			&=&\frac{\pi a}{2}\left(\sqrt{\frac{\left[B+(a+n\hbar)^2\sqrt{\lambda}\right]^2}{a^2(a+n\hbar)^2\lambda}}+\sqrt{\frac{\left[B-(a+n\hbar)^2\sqrt{\lambda}\right]^2}{a^2(a+n\hbar)^2\lambda}}-2 \right)                  \nonumber\\
		&=&\frac{\pi a}{2}\left[\frac{B+(a+n\hbar)^2\sqrt{\lambda}-B+(a+n\hbar)^2\sqrt\lambda - 2 a(a+n\hbar)\sqrt{\lambda}}{a(a+n\hbar)\sqrt{\lambda}}\right]\nonumber\\
		&=&n\pi\hbar~.
		\label{eq.class2b1}\end{eqnarray}
	
\medskip
\noindent{\bf Case IIB3: $\lambda>0$ and $a>0$. }
	
	For this case, $\varepsilon_1=1$ and $\varepsilon_2=1,$ and $0<a^4\lambda<B^2$. We note that $(y_L \, + y_R)^2=\frac{4B^2}{a^4\lambda}>4$. Since $y^2$ cannot cross 1 within the domain, this means either that $1<y_L<y_R$ or $y_L<y_R<-1$. Therefore $y'>0$ and $\varepsilon_3=1.$ 
		Therefore, 
	\begin{eqnarray}
		I_{S}&=&  a \int_{y_L}^{y_R}\frac{\sqrt{\left(y_R-y \right)\left( y-y_L \right)}}{y^2-1}\,dy = a I_{5}~,
	\end{eqnarray}
	where $I_5$ is $I_{5a}$ from Appendix \ref{AppendixC} when $1<y_L<y_R$, and $I_{5b}$ when $y_L<y_R<-1$.
		Either way, this leads to 
	 \begin{eqnarray}
	 	I_{S}
	 	&=&\frac{\pi a}{2}\left[\frac{-B+(a+n\hbar)^2\sqrt{\lambda}+B+(a+n\hbar)^2\sqrt\lambda - 2 a(a+n\hbar)\sqrt{\lambda}}{a(a+n\hbar)\sqrt{\lambda}}\right]\nonumber\\
	 	&=&n\pi\hbar.
	 	\label{eq.class2b2}\end{eqnarray}

\subsection{Class III Potentials: Neither $f_1$ nor $f_2$ are constant}

For Class III potentials, neither $f_1$ nor $f_2$ are constant. They must be linearly independent of each other, otherwise they would be equivalent to Class II. Therefore, we have  $W=a f_1+f_2+u$. We differentiate  Eq. (\ref{eq.PDE1}) twice with respect to $a$, obtaining 
$
	f_1\left(3\ddot{u}+a\dddot{u} \right)+ f_2 \dddot{u} = - 3\dot{u}\ddot{u} - u\dddot{u} - \dddot{g}/2.
$
Since $f_1$ and $f_2$ are linearly independent, $f_2 \dddot{u}=0$, which means $\dddot{u}=0$. Thus $f_1\left(3\ddot{u}+a\dddot{u} \right)=0$, so $\ddot{u}=0$. Therefore, $u(a)=\mu a + \nu$ for constants $\mu$ and $\nu$. We can regroup terms to include $\mu$ in $f_1$ and $\nu$ in $f_2$, leaving us with 
\begin{equation}
	W(x,a)=a f_1(x) + f_2(x).
\end{equation}
Returning to Eq.~(\ref{eq.PDE1}), we now have 
$
	\left(f_1^2 - f_1'\right)a + f_1 f_2 -f_2'+\dot{g}/2=0.
$
Grouping terms by powers in $a$, each power must separately reduce. We therefore have
\begin{eqnarray}
	f_1'&=&f_1^2-\lambda,\\
	f_2'&=&f_1 f_2 +\omega/2,\\
	\dot{g}&=&-2 a \lambda + \omega,
\end{eqnarray}
for constants $\lambda$ and $\omega$.

We define Subclass IIIA, where $\lambda=0$, and Subclass IIIB, where $\lambda\neq0$. 

\subsubsection{Subclass IIIA: $W=a f_1+f_2$; $f_1'=f_1^2$}\label{sec:ClassIIIA}

Because $\lambda=0$, we have $f_1'=f_1^2$, so $f_1\neq0$ and $f_1'\neq0$ within the domain. We also have $f_2'=f_1 f_2 +\omega/2$. These are solved by functions of the form 
$
	f_2 = \alpha f_1 - \frac{\omega}{4f_1}.
$
Consequently $W=(a+\alpha) f_1 - \frac{\omega}{4f_1}$. With a redefinition of \(a\), this leads to
 \begin{equation}
  	W= a f_1 - \frac{\omega}{4f_1}.\label{eq:ClassIIIAW}
 \end{equation}
The SWKB integral becomes
 \begin{equation*}
	I_{S}=  \int_{x_L}^{x_R} \sqrt{E_n-\left(a f_1-\frac{\omega}{4f_1}\right)^2}\, {\rm d}x 
	=\int_{x_L}^{x_R} \frac{\sqrt{E_n f_1^2-\left(af_1^2-\frac{\omega}{4}\right)^2}}{\varepsilon_1f_1}\, {\rm d}x, 
\end{equation*}
where $\varepsilon_1=1$ when $f_1>0$, and $\varepsilon_1=-1$ when $f_1<0$.
We change integration variable by defining $y\equiv f_1^2$. In this case, $y'=2f_1^3$, so $y'$ has the same sign as $f_1$, and the SWKB integral becomes 
\begin{equation}
	I_{S} = \int_{y_L}^{y_R} \frac{\sqrt{E_n y- \left(a y - \omega/4\right)^2}}{2y^2}\, {\rm d}y, 
\end{equation}
where 
\begin{equation*}
	y_L=\frac{2 E_n +a\omega-2\sqrt{E_n}\sqrt{E_n+a\omega}}{4a^2}~~,
\qquad
	y_
	R=\frac{2 E_n +a\omega+2\sqrt{E_n}\sqrt{E_n+a\omega}}{4a^2}.
\end{equation*}
The integral can be written as
 \begin{equation*}
	I_{S}= \frac{a}{2}\int_{y_L}^{y_R} \frac{\sqrt{(y_R-y)(y-y_L)}}{2y^2}\, {\rm d}y =\frac a4 \,I_{2B},
\end{equation*}
where $I_{2B}$ is given in Appendix \ref{AppendixC}. Thus 
 \begin{eqnarray}
	I_{S}&=& \frac{a\pi\left(y_L+y_R-2\sqrt{y_L y_R}\right)}{4\sqrt{y_L y_R}}. \nonumber 
\end{eqnarray}
Substituting the integration limits  gives
 \begin{eqnarray}
	I_{S}&=& \frac{\pi}{|\omega|}\left(E_n + a\omega/2 - |a \omega|/2\right).\label{eq:IIIAsoln}
\end{eqnarray}

We note the following for the broken and unbroken SUSY phases:
\begin{itemize}
	\item \textbf{Unbroken SUSY:} Since $f_1$ cannot change sign, from Eq. (\ref{eq:ClassIIIAW}) $a$ and $\omega$ must have the same sign. Because $\dot{g}=\omega>0$, this requires $a>0$. 
	\item \textbf{Broken SUSY:} SUSY is broken when $\omega$ and $a$ have opposite signs. 
\end{itemize}

In the unbroken SUSY phase, $|a|=a$, and the energies are given by $E_n= \omega n \hbar.$
Therefore, from Eq.~(\ref{eq:IIIAsoln}):
 \begin{eqnarray}
	I_{S}&=& n\pi\hbar.
\end{eqnarray}

\subsubsection{Subclass IIIB: $W=a f_1+f_2$; $f_1'=f_1^2-\lambda$; $\lambda\neq 0$}

In this case, we have $f_1'=f_1^2-\lambda$, where $\lambda\neq 0$. Therefore, $f_1^2\neq\lambda$ and $f_1'\neq0$ within the domain. We also have $f_2'=f_1 f_2 +\omega/2$, which leads to
\begin{equation}
f_2 = \frac{-f_1\omega}{2\lambda}+B\sqrt{\left|f_1^2-\lambda\right|},
\end{equation}
where $B$, $\omega$, and $\lambda$ are all real constants. 
We then have $W=(a-\frac{\omega}{2\lambda}) f_1 +B\sqrt{\left|f_1^2-\lambda\right|}$. We redefine $ a-\frac{\omega}{2\lambda}$ as \(a\), which leads to
\begin{equation}
	W= a f_1 +B\sqrt{\left|f_1^2-\lambda\right|}.
\end{equation}
We note that $\dot{g}=-2 a\lambda  >0$, so $a \lambda<0$. For unbroken SUSY,
\begin{equation}
	E_n=-2n\hbar\lambda a -n^2\hbar^2\lambda.\label{eq:ClassIIISpectrum}
\end{equation}
We now consider three cases.

\medskip
\noindent{\bf Case IIIB1: $\lambda<0$}\label{sec:ClassIIIB1}

For notational convenience, we define $\alpha\equiv-\lambda>0$, so $f_1'=f_1^2+\alpha>0$.
We define the complex function
\begin{equation}
	y\equiv\frac{\sqrt{\lambda}-f_1}{\sqrt{\lambda}+f_1}=\frac{i \sqrt{\alpha}-f_1}{i \sqrt{\alpha}+f_1}.
\end{equation}
Because \(f_1\) is real, $y\neq 0$. We now define two real functions,
\begin{equation}
{\cal S}\equiv\frac{y^{1/2}-y^{-1/2}}{2 i}~,\qquad
{\cal C}\equiv\frac{y^{1/2}+y^{-1/2}}{2}.
\end{equation}
With these definitions, 
\begin{equation}
	{\cal C}^2 + 	{\cal S}^2 = 1, \qquad {\cal S}'=\sqrt{\alpha}~{\cal C}, \qquad {\rm and} \qquad {\cal C}'=-\sqrt{\alpha}~{\cal S},
\end{equation}
where $-1<{\cal S}<1$, and $0<|{\cal C}|<1$.

We  can rewrite $f_1$ in terms of ${\cal S}$ and ${\cal C}$:
\begin{equation}
f_1=\sqrt{\alpha}\,\frac{{\cal S}}{{\cal C}}.
\end{equation}
With this identification, we can write 
\begin{equation}
	W=\frac{\sqrt{\alpha}}{\cal C}\left(a\,{\cal S}+\varepsilon_1B\right),
\end{equation}
where $\varepsilon_1=1$ when ${\cal C}>0$, and $\varepsilon_1=-1$ when ${\cal C}<0$. 

Recall ${\cal S}'=\sqrt{\alpha}\, {\cal C}$, which has the same sign as $\epsilon_1$. Also, ${\cal C}^2= 1- {\cal S}^2$, so 
 \begin{eqnarray}
	I_{S}&=& 
	 \varepsilon_1 \int_{{\cal S}_L}^{{\cal S}_R} \frac{\sqrt{E_n-\left( \sqrt{\alpha}/{\cal C}\right)^2\left(a{\cal S} + \varepsilon_1 B\right)^2}}{{\cal S'}}{\rm d}{\cal S}\nonumber \\
	&=&\frac{1}{\sqrt{\alpha}} \int_{{\cal S}_L}^{{\cal S}_R} \frac{\sqrt{E_n ({1-\cal S}^2)-\alpha  \left(a{\cal S} + \varepsilon_1 B\right)^2}}{1-{\cal S}^2}{\rm d}{\cal S},
\end{eqnarray}
where the integration limits are 
\begin{equation}
	{\cal S}_L=\frac{-a \alpha \varepsilon_1 B-\sqrt{E_n^2+a^2 E_n \alpha -E_n\alpha B^2}}{E_n+a^2\alpha},
\end{equation}
and
\begin{equation}
	{\cal S}_R=\frac{-a \alpha \varepsilon_1 B+\sqrt{E_n^2+a^2 E_n \alpha -E_n\alpha B^2}}{E_n+a^2\alpha}.
\end{equation}
This gives
\begin{equation*}
	I_{S} = \frac{\sqrt{E_n+a^2\alpha}}{\sqrt{\alpha}}\int_{
		{\cal S}_L}^{{\cal S}_R} 
		\frac{\sqrt{({\cal S}_R -{\cal S})({\cal S} -{\cal S}_L)}}{1-{\cal S}^2} {\rm d}{\cal S} =
	\frac{\sqrt{E_n+a^2\alpha}}{\sqrt{\alpha}}\,I_4,
\end{equation*}
where $-1 < {\cal S}_L <{\cal S}_R< 1$ and $I_4$ is given in  Appendix \ref{AppendixC}. Thus 
 \begin{eqnarray}
	I_{S} 
	 &=& \frac{\pi\sqrt{E_n+a^2\alpha}}{2\sqrt{\alpha}}\left[2-\sqrt{\frac{\alpha(a+\varepsilon_1 B)^2}{E_n+a^2\alpha}}-\sqrt{\frac{\alpha(a-\varepsilon_1 B)^2}{E_n+a^2\alpha}}\right] \nonumber\\
	 &=& 	\frac{\pi}{2\sqrt{\alpha}}\left[2\sqrt{E_n+a^2\alpha} - \sqrt{\alpha}\left|a+\varepsilon_1 B\right| - \sqrt{\alpha}\left|a-\varepsilon_1 B\right|
\right].
\label{eq:IIIB1integral}\end{eqnarray}

We note the following conditions for the broken vs. unbroken SUSY phases:
\begin{itemize}
	\item \textbf{Unbroken SUSY:} We know 	$W=\frac{\sqrt{\alpha}}{\cal C}\left(a{\cal S}+\varepsilon_1B\right),$ $\varepsilon_1 B$ is a constant, ${\cal C}$ cannot change sign, and $-1<{\cal S}<1$. Therefore, unbroken SUSY requires $|a|>|B|$. 
	\item \textbf{Broken SUSY:}
	For the SWKB integral to have two turning points, \(|a|\ne|B|\).
	From the argument above, broken SUSY requires $|a|<|B|$. 
\end{itemize}

For unbroken SUSY, $|a|>|B|$. Since $\dot{g}>0$,  $a>0$, so $a\pm\varepsilon_1 B>0$. Using this  together with the energy spectrum from Eq.~(\ref{eq:ClassIIISpectrum}), the integral simplifies to 
 \begin{eqnarray}
	I_{S}
	&=&\pi\left[\sqrt{(a+n\hbar)^2}-a\right] = n\pi\hbar.
\end{eqnarray}

\noindent{\bf Case IIIB2: $0\leq f_1^2<\lambda$ }\label{Sec:classIIIB2}

We define the real function
\begin{equation}
	y\equiv\frac{\sqrt{\lambda}-f_1}{\sqrt{\lambda}+f_1}.
\end{equation}
Noting that $0<y<\infty$ and  $y'=2 \sqrt{\lambda}\, y$, we introduce two real functions
\begin{equation}
	{\cal S}\equiv\frac{y^{1/2}-y^{-1/2}}{2} ~~, \quad {\cal C}\equiv\frac{y^{1/2}+y^{-1/2}}{2}.
\end{equation}
With these definitions, 
\begin{equation}
	{\cal C}^2 - {\cal S}^2 = 1, \qquad {\cal S}'=\sqrt{\lambda}{\cal C}, \qquad {\rm and}\qquad {\cal C}'=\sqrt{\lambda}{\cal S},
\end{equation}
where $0<{\cal C}<\infty$, and $-\infty<|{\cal S}|<\infty$.
Therefore,
\begin{equation}
	f_1=-\sqrt{\lambda}\,\frac{{\cal S}}{{\cal C}}~~,\quad
	W=\frac{\sqrt{\lambda}}{\cal C}\left(-a\,{\cal S}+B\right).
\end{equation}

We note the following conditions for the broken vs. unbroken SUSY phases:
\begin{itemize}
	\item \textbf{Unbroken SUSY:} Since $B$ is a constant and $-\infty<{\cal S}<\infty$, SUSY is unbroken for all values of the parameters. Since $a\lambda<0$, it follows that $a<0$. 
	\item \textbf{Broken SUSY:}
	From the argument above, this superpotential does not enter the broken SUSY phase.
\end{itemize}

 Recall ${\cal S}'=\sqrt{\lambda}\, {\cal C}>0$. We again change integration variable:
\begin{eqnarray}
	I_{S}&=& 
	 \int_{{\cal S}_L}^{{\cal S}_R} \frac{\sqrt{E_n-\lambda/{\cal C}^2\left(-a\,{\cal S} + B\right)^2}}{{\cal S'}}\,{\rm d}{\cal S}\nonumber \\
	&=&\frac{1}{\sqrt{\lambda}} \int_{{\cal S}_L}^{{\cal S}_R} \frac{\sqrt{E_n ({1+\cal S}^2)-\lambda  \left(-a{\cal S} + B\right)^2}}{1+{\cal S}^2}\,{\rm d}{\cal S},
\end{eqnarray}
where the integration limits are 
\begin{equation}
	{\cal S}_L=\frac{a \lambda B-\varepsilon_2\sqrt{-E_n^2+a^2 E_n \lambda +E_n\lambda B^2}}{-E_n+a^2\lambda},
\end{equation}
\begin{equation}
	{\cal S}_R=\frac{a \lambda B+\varepsilon_2\sqrt{-E_n^2+a^2 E_n \lambda +E_n\lambda B^2}}{-E_n+a^2\lambda},
\end{equation}
where $\varepsilon_2=1$ if $a^2\lambda > E_n$ and $\varepsilon_2=-1$ if $a^2\lambda < E_n$.
Therefore,
\begin{equation*}
	I_{S} = \frac{\sqrt{a^2\lambda-E_n}}{\sqrt{\lambda}}
	\int_{{\cal S}_L}^{{\cal S}_R} 
	\frac{\sqrt{({\cal S}_R -{\cal S})({\cal S} -{\cal S}_L)}}{1+{\cal S}^2} {\rm d}{\cal S} 
	= \frac{\sqrt{a^2\lambda-E_n}}{\sqrt{\lambda}}\,I_3
\end{equation*}
where $I_3$ is given in  Appendix \ref{AppendixC}. So 
\begin{equation}
	I_{S} = \frac{\sqrt{-E_n+a^2\lambda}}{\sqrt{\lambda}}\left[\frac{\pi}{\sqrt{2}}\left(\sqrt{(1+{\cal S}_L^2)(1+{\cal S}_R^2)}-{\cal S}_L {\cal S}_R + 1\right)^{1/2}-\pi\right]
	= n\pi\hbar.
\end{equation}

\medskip
\noindent{\bf Case IIIB3: $0<\lambda<f_1^2$}\label{Sec:classIIIB3}

\noindent
We define $\varepsilon_1=1$ if $f_1>\sqrt{\lambda}$ and $\varepsilon_1=-1$ if $f_1<-\sqrt{\lambda}$, together with 
\begin{equation}
	y\equiv\frac{f_1+\varepsilon_1\sqrt{\lambda}}{f_1-\varepsilon_1\sqrt{\lambda}}.
\end{equation}
Then $y>1$, and $y'=-2\sqrt{\lambda}\,  \varepsilon_1 y$, which has fixed sign, opposite to that of $f_1$. We now define two real functions of $y$:
\begin{equation}
	{\cal S}\equiv\frac{y^{1/2}-y^{-1/2}}{2}
~~,\quad
	{\cal C}\equiv\frac{y^{1/2}+y^{-1/2}}{2}.
\end{equation}
Then $1<{\cal C}<\infty$,  $0<{\cal S}<\infty$, and 
\begin{equation}
	{\cal C}^2 - {\cal S}^2 = 1~, \qquad {\cal S}'=-\varepsilon_1\sqrt{\lambda}\,{\cal C}~, \qquad {\cal C}'=-\varepsilon_1\sqrt{\lambda}{\cal S}.
\end{equation}
Therefore
\begin{equation}
	\label{eq:IIIB3}
	f_1=\varepsilon_1\sqrt{\lambda}\,\frac{{\cal C}}{{\cal S}}~~,\qquad
	W=\frac{\sqrt{\lambda}}{\cal S}\left(\varepsilon_1 a\,{\cal C}+B\right).
\end{equation}
We obtain
\begin{eqnarray}
	I_{S}
	&=&\frac{1}{\sqrt{\lambda}} \int_{{\cal C}_L}^{{\cal C}_R} \frac{\sqrt{E_n ({\cal C}^2-1)-\lambda  \left(\varepsilon_1 a{\cal C} + B\right)^2}}{{\cal C}^2-1}{\rm d}{\cal C},
\end{eqnarray}
where the integration limits are
\begin{equation}
	{\cal C}_L=\frac{-\varepsilon_1 a \lambda B-\varepsilon_2\sqrt{E_n^2-a^2 E_n \lambda +E_n\lambda B^2}}{-E_n+a^2\lambda},
\end{equation}
\begin{equation}
	{\cal C}_R=\frac{-\varepsilon_1 a \lambda B+\varepsilon_2\sqrt{E_n^2-a^2 E_n \lambda +E_n\lambda B^2}}{-E_n+a^2\lambda},
\end{equation}
where $\varepsilon_2=1$ if $a^2\lambda > E_n$ and $\varepsilon_2=-1$ if $a^2\lambda < E_n$.

Then the integral becomes
\begin{equation}
	I_{S} = \frac{\sqrt{-E_n+a^2\lambda}}{\sqrt{\lambda}}
	\int_{{\cal C}_L}^{{\cal C}_R} 
	\frac{\sqrt{({\cal C}_R -{\cal C})({\cal C} -{\cal C}_L)}}{{\cal C}^2-1}\, {\rm d}{\cal C}
	=\frac{\sqrt{-E_n+a^2\lambda}}{\sqrt{\lambda}}\, I_{5a},
\end{equation}
where $1 < {\cal C}_L < {\cal C}_R < \infty$ and $I_{5a}$ is given in  Appendix \ref{AppendixC}. So 
 \begin{eqnarray}
 	I_{S} 
	&=& \frac{\pi\sqrt{a^2\lambda-E_n}}{2\sqrt{\lambda}}\left[\sqrt{\frac{\lambda(\varepsilon_1 B-a)^2}{a^2\lambda-E_n}}-\sqrt{\frac{\lambda(\varepsilon_1 B+a)^2}{a^2\lambda-E_n}}-2\sqrt{a^2\lambda-E_n}\right] \nonumber\\
	&=& 	
	\frac{\pi}{2\sqrt{\lambda}}\left[
\sqrt{\lambda}\left|(\varepsilon_1 B-a)\right|-
\sqrt{\lambda}\left|(\varepsilon_1 B+a)\right|-
2\left|a^2\lambda-E_n\right|\right].
	\label{eq:IIIB3integral}\end{eqnarray}

We note the following conditions for the broken vs. unbroken SUSY phases:
\begin{itemize}
	\item \textbf{Unbroken SUSY:} 
	 We know 	$W=\frac{\sqrt{\lambda}}{\cal S}\left(\varepsilon_1 a \, {\cal C}+B\right)$, $\varepsilon_1 a$ is a constant, ${\cal S}>0$, and ${\cal C}>1$. Therefore, $W$ has opposite sign at the two ends of its domain if $\varepsilon_1a$ and $B$ have opposite signs and $|B|>|a|$.
	\item \textbf{Broken SUSY:}
	From the argument above, SUSY is broken if either $\varepsilon_1 a$ has the same sign as $B$, or if  $|a|>|B|$.
\end{itemize}

For the unbroken phase, the SWKB integral simplifies to 
\begin{equation}
	I_{S} = \frac{\pi}{2}\sqrt{(a+n\hbar)^2}\left[\sqrt{\frac{(-a^2+B^2)^2}{(a+n\hbar)^2}}-\sqrt{\frac{(a^2+B^2)^2}{(a+n\hbar)^2}}-2\right] =  n\pi\hbar.
\end{equation}

By considering the three classes and their associated subclasses, we have shown that in each case, the SWKB integral $I_S=n\pi\hbar$. Therefore, SWKB is exact for all conventional superpotentials in their unbroken phase, as a direct consequence of their additive shape invariance.

\section{SWKB Exactness for Conventional Potentials in the Broken Phase}\label{sec:BSWKB}

As we showed in Eq. (\ref{eq:bswkb}), for broken SUSY, the BSWKB quantization condition is
\begin{equation*}
	I_S \equiv 
\int_{x_L}^{x_R} \sqrt{E_n(a)-W^2(x,a)} ~dx  = \left( n+\frac12\right)  \pi\hbar~, \quad \mbox{where}~ n = 0,1,2,\cdots .
\end{equation*}
Recently, the authors of Ref.~\cite{Gangopadhyaya2021} proved that this equation gives exact eigenenergies  for conventional superpotentials for which the integral is well-defined. We review these results, showing that they follow directly from the results in Sec.~\ref{sec:UBSWKB} due to the existence of a discrete shape invariance that leads to a change of phase between broken and unbroken supersymmetry. 

\begin{figure}[h!]
	\centering
	\includegraphics[width=0.6\linewidth]{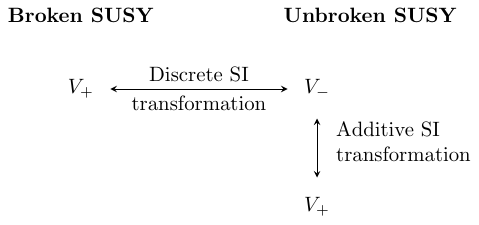}
	\caption{Schematic illustrating the shape invariant transformations used to determine the spectra for the systems in their broken phases. Reprinted with permission from Gangopadhyaya, et al.\cite{Gangopadhyaya2021} \copyright 2021 IOP Publishing Ltd.}
	\label{fig:transformations}
\end{figure}

Following Refs.~\cite{Gangopadhyaya2001,Gangopadhyaya2021} , we start with a discrete change of parameter that takes the system from the broken to an associated unbroken phase. We use the known spectrum in the unbroken phase and the isospectral relations to determine the eigenspectrum for the broken phase, as illustrated by Fig. \ref{fig:transformations}.
We consider each of the three conventional classes identified in Sec. \ref{sec:conventional}, and show that $I_S=(n+1/2)\pi\hbar$ for all cases in which the integral is properly defined. Since these classes are exhaustive, we thus prove that the BSWKB condition is exact for all such cases.

\subsection{Class I and Class II}
As discussed in Secs. \ref{Sec:classISWKB}-\ref{Sec:classIISWKB}, there cannot be two roots of $E_n-W^2$ in the broken phase, so the BWSKB condition does not apply to these classes.

\subsection{Class III}
For Class III, $W=a f_1+ f_2$. 
As we show below, $W$ can have an extremum, in which case $W^2=E_n$ can have two intersection points. 
Here $f_1^\prime=f_1^2 -\lambda$ and $f_2^\prime = f_1 f_2 +\omega/2$, where $\lambda$ and $\omega$ are constants; $\lambda= 0$ for Subclass IIIA and   $\lambda\neq0$ for Subclass IIIB.

\subsubsection{Subclass IIIA: $\lambda =0$}

In the broken phase, $\omega$ and $a$ have opposite signs; they have the same sign in the unbroken phase. Therefore, the discrete transformation $a \to -a$ changes between broken and unbroken SUSY. 
From Eq. (\ref{eq:ClassIIIAW}),
\begin{equation*}
	W= a f_1 - \frac{\omega}{4f_1}.
\end{equation*}
The partner potentials are given by
\begin{equation}
	V_\pm = a (a\pm \hbar) f_1^2 \, + \,\frac1{16} \frac{\omega^2}{f_1^2}  \, - \, \frac1{2} a \, \omega \, \pm \,
	\frac14\hbar \, \omega~.
\end{equation}
They are shape invariant under two different parameter changes:
\begin{itemize}
	\item Additive Shape Invariance:  $V_+(x,a) - V_-(x,a+\hbar) =  \hbar \omega $ ,
	\item Discrete Shape Invariance: $V_+(x,a) - V_-(x,-a) = -(2 a - \hbar)\, \omega/2 $ .
\end{itemize}
Since the discrete shape invariance transforms the broken into the unbroken phase and the energy for the unbroken phase is $g(a+n\hbar) - g(a)= \hbar n \omega$,  the energy in the broken phase is given by
$E_n= n \hbar \omega -(2 a - \hbar)\, \omega/2= \left[\hbar(n+ 1/2) - a\right]\,\omega$.

For both the broken and unbroken phases, we showed in Eq. (\ref{eq:IIIAsoln})  that  
\begin{equation*}
I_{S}= \frac{\pi}{|\omega|}\left(E_n + a\,\omega/2 - |a\, \omega|/2\right) .
\end{equation*}
In the broken phase, $E_n= [\hbar(n+ 1/2) - a]\,\omega$, and $|a\, \omega|=-a \,\omega,$ so this reduces to
\begin{equation}
	I_{S}= \pi\hbar\,(n+ 1/2).
\end{equation}

\subsubsection{Subclass IIIB: $\lambda \ne 0$}
We consider Cases IIIB1, IIIB2, and IIIB3 separately. 

\medskip
\noindent{\bf Case IIIB1: $\lambda<0$}\label{Sec:B-classIIIB1}

Recall that in Sec. \ref{sec:ClassIIIB1}  we defined  $\alpha\equiv-\lambda>0$. In this case, $W=a f_1 + B{f}_2$, where ${f}_2=\sqrt{f_1^2+\alpha}$. The partner potentials become
\begin{eqnarray}V_\pm ({a},B,x) &=&
	\left[ {a}({a}\pm \hbar) + B^2\right] \, {f_1}^2
	+2B  \left( {a} \pm \hbar/2 \right)  f_1 f_2
	+\alpha  \left(B^2 \pm {a} \hbar\right).
\end{eqnarray}
In addition to additive shape invariance generated by $a\rightarrow a+\hbar$, these two potentials also satisfy a discrete shape invariance via $a \rightarrow B+\hbar/2, B\rightarrow a+\hbar/2$. That is,
\begin{itemize}
	\item Additive Shape Invariance:  $V_+({a},B,x) - V_-({a}+\hbar,B,x)  =  \alpha\left[({a}+\hbar)^2-{a}^2\right]$ ,
	\item Discrete Shape Invariance: $V_+({a},B,r) - V_-(B+\frac{\hbar}{2}, {a}+\frac{\hbar}{2},r) =   {\alpha  \left[(B +\frac{\hbar}{2})^2-{a}^2\right]}$. 
\end{itemize}
The discrete shape invariance changes between broken and unbroken phases.\footnote{
	We note that $V_+({a},B)= V_-(-{a},-B)$, so we can choose the sign of $B$ without loss of generality by switching the labels of $V_+$ and $V_-$. We choose \(B>0\). Since the unbroken phase corresponds to $a >|B|$  and the broken phase corresponds to  $|{a}|<|B|$, the discrete shape invariance corresponds to a transition from the broken phase to the unbroken phase.}

We now find the spectrum in the broken phase from the corresponding unbroken phase. We recall that for unbroken SUSY, the spectrum is given in Eq.~(\ref{eq:ClassIIISpectrum}) as $E_n=-2n\hbar\lambda{a}-n^2\hbar^2\lambda$.
Let $E_n^B$ be the energy for the system with potential $V_+ (a,B,r)$ with broken SUSY, and  $E_n$ be the energy for $V_-(B+\hbar/2, a+\hbar/2,r)$ with unbroken SUSY. Then from the discrete shape invariance condition we have 
\begin{eqnarray}E_n^B&=&E_n+ {\alpha  \left[\left(B +\frac{\hbar}{2}\right)^2-{a}^2\right]}=\alpha  \left[{\left(B +n\,\hbar+\frac{\hbar}{2}\right)^2}-{a}^2\right].
\end{eqnarray}
We recall from Eq.~(\ref{eq:IIIB1integral})
that for Class IIIB,
\begin{equation}
 I_{S}	=	\frac{\pi}{2\sqrt{\alpha}}\left[2\sqrt{E_n^B+{a}^2\alpha} - \sqrt{\alpha}\left|{a}+\varepsilon_1 B\right| - \sqrt{\alpha}\left|{a}-\varepsilon_1 B\right|
\right].
\end{equation}
For the broken phase, $B>|{a}|$, so this yields
\begin{equation}
	I_{S}	=	\pi\big[(B +n\,\hbar+\hbar/2) - B
	\big]
	=	\pi\hbar(n+1/2).
\end{equation}

\bigskip

\noindent
\textbf{\bf Case IIIB2:} $\lambda<0, \lambda>f_1^2$ \label{Sec:B-classIIIB2}

\noindent
As discussed in Sec. \ref{Sec:classIIIB2}, the superpotential does not enter a broken SUSY phase.

\bigskip

\noindent
\textbf{\bf Case IIIB3:} $0<\lambda<f_1^2$

\noindent 
In this case \(	W=\frac{\sqrt{\lambda}}{\cal S}\left(\varepsilon_1 a\,{\cal C}+B\right)\) from Eq. (\ref{eq:IIIB3}).
As discussed in Sec. \ref{Sec:classIIIB3}, SUSY is broken if $\varepsilon_1 {a}B <0$ and/or if  $|{a}|>|B|$.
If $\varepsilon_1 {a}B <0$, $W'$ is monotonic in $x$, so the broken phase does not have two integration limits.
If $\varepsilon_1 {a}B >0$, and $|{a}|>|B|$, we can follow the derivation of energy in the broken phase from Class IIIB1, replacing $\alpha$ with $-\lambda$. We get

\begin{equation}
	E_n^B=E_n+ {\alpha  \left[\left(B +\frac{\hbar}{2}\right)^2-{a}^2\right]}
	= \lambda  \left[{a}^2-{\left(B +n\,\hbar+\frac{\hbar}{2}\right)^2}\right].
\end{equation}
To ensure proper ordering of energy levels, we require $B +n\,\hbar+\frac{\hbar}{2}<0$ for all bound states $E_n$.
We recall from Eq.~(\ref{eq:IIIB3integral}) that for Case IIIB3,
\begin{equation}
	I_{S}	 = 	
\frac{\pi\sqrt{{a}^2\lambda-E_n}}{2\sqrt{\lambda}}\left[
\sqrt{\lambda}\left|(\varepsilon_1 B-{a})\right|-
\sqrt{\lambda}\left|(\varepsilon_1 B+{a})\right|-
2\sqrt{{a}^2\lambda-E_n}\right].
\end{equation}
This yields
\begin{equation}
	I_{S}	=	 \pi\left(\left|B\right|-\left|B +n\,\hbar+\frac{\hbar}{2}\right| \right) 
			= \pi \hbar (n + 1/2).
\end{equation}

\begin{figure}
	\centering
	\includegraphics[width=0.9\linewidth]{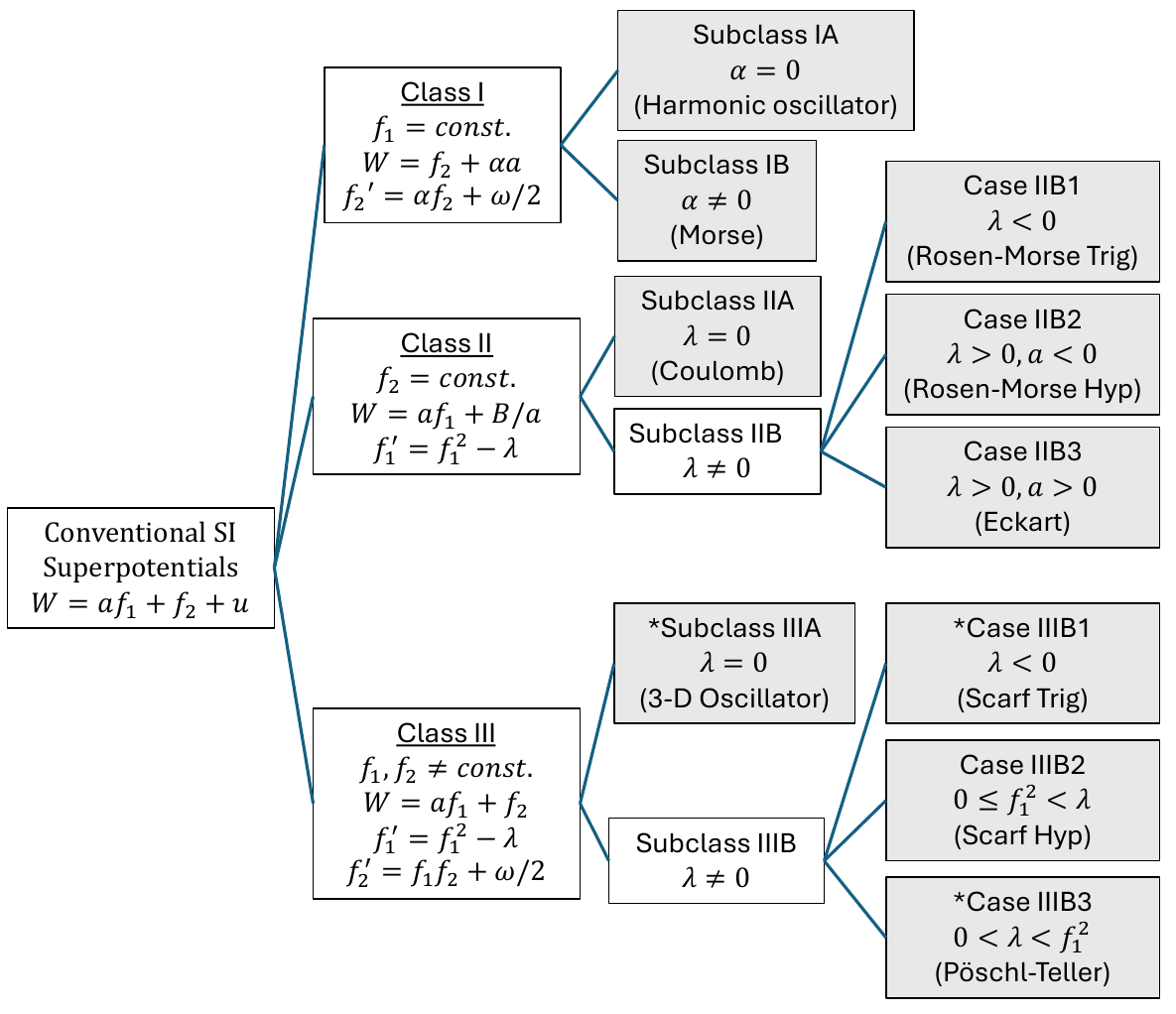}
	\caption{Categorization of conventional shape invariant superpotentials. While SWKB is exact for all superpotentials in the unbroken phase, for the broken phase the  only instances where the SWKB integral is properly defined are marked with asterisks.}
	\label{fig:additivesi}
\end{figure}

We conclude this section with Fig. \ref{fig:additivesi} which summarizes the exhaustive list of conventional shape invariant superpotentials. 

\section{Generalized Langer Correction}\label{sec:langer}

Recent work \cite{Gangopadhyaya2023} demonstrated that a universal correction renders the WKB quantization condition exact for all conventional SIPs in their unbroken phase, and that this correction follows from the exactness of SWKB described in Sec. \ref{sec:UBSWKB}. We review this result here.

As discussed in Sec.~\ref{sec:semiclassical}, the WKB quantization condition given by
\(\int_{x_L}^{x_R} \sqrt{E_n-{V}(x) } ~dx = \left( n+\frac12 \right) \pi \hbar~\) yields exact energies for the cases of the one-dimensional harmonic oscillator and Morse potentials, which correspond to Class I SIPs. However, the energy spectrum is not exact for SIPs of other classes.

We will now introduce a generalized term $\Delta V(x)$ such that the  corrected potential $\tilde{V}(x) \equiv V_-(x)+\Delta V(x)$ yields exact eigenenergies for all conventional SIPs: 
\begin{equation}
	\int_{x_L}^{x_R} \sqrt{E_n-\tilde{V}(x) } ~dx = \left( n+\frac12 \right) \pi \hbar~. 
\end{equation}
In Sec.~\ref{sec:UBSWKB}, we showed that the SWKB quantization condition yields exact eigenvalues for all conventional SIPs in the unbroken supersymmetric phase:
\begin{equation}
	\int_{x_L}^{x_R} \sqrt{E_n-W^2(x) } ~dx = n \pi \hbar~. \label{eq.SWKB1a}
\end{equation}

We begin our proof by observing that we can treat the quantum number $n$ in Eq. (\ref{eq.SWKB1a}) as a parameter; i.e., the condition is exact for any value of $n$, provided $E_n-W^2(x)$ has two roots. In particular, for $n\to n+1/2$, the condition becomes
\begin{equation}
	\int_{x_L}^{x_R} \sqrt{E_{n+1/2}-W^2(x) } ~dx = \left( n+\frac12\right)  \pi \hbar~. \label{eq.SWKB11}
\end{equation}
Integrating Eq. (\ref{eq.PDE1}) we obtain:
\begin{equation}
	\label{eq:VL-SWKB1}
	\int_{a-\hbar/2}^a \frac{\partial}{\partial a} \Big( W^2(x,a) + g(a) \Big) da = 2 \int_{a-\hbar/2}^a \frac{\partial }{\partial x} \,W(x,a)\, da~,
\end{equation}
which, along with Eq. (\ref{eq.PDE2solution}), leads to 
\begin{equation}
	\label{eq:VL-SWKB2}
	W^2(x,a) - \hbar \Big(af_1'(x) +f_2'(x) \Big) + \frac 14\, \hbar^2 f_1'(x) + g(a)
	= W^2(x,a-\hbar/2) + g(a-\hbar/2)~.
\end{equation}
Subtracting $g(a+n\hbar)$ from both sides of Eq. (\ref{eq:VL-SWKB2}) and using Eq. (\ref{eq:En}), we get
\begin{equation}
	\label{eq:VL-SWKB3}
	E_n(a) - \left( V_-(x,a) +  \frac 14 \,\hbar^2 f_1'(x) \right) 
	= E_{n+ \frac 12}(\tilde{a}) - W^2(x,\tilde{a}) ~,
\end{equation}
where $\tilde{a} \equiv a-\hbar/2$ and 
\[
E_{n+ \frac 12}(\tilde{a}) = g(\tilde{a}+(n+\tfrac{1}{2})\hbar) - g(\tilde{a}) .
\]
Eq. (\ref{eq:VL-SWKB3}) suggests the following generalized Langer correction
\begin{equation}
	\Delta V \equiv \frac 14 \,\hbar^2 f_1'(x).\label{eq.langercorrection}
	\end{equation}
Because $\tilde{V}(x)=V_-(x)+\Delta V(x)$, we have arrived at the main result:
\begin{equation}
	\int_{x_L}^{x_R} \sqrt{E_n(a) - \tilde{V}(x,a) }~dx =
	\int_{x_L}^{x_R} \sqrt{E_{n+ \frac 12}(\tilde{a}) - W^2(x,\tilde{a})} ~dx 
		=
	 \left( n+ \frac 12\right)  \pi \hbar. \label{eq.SWKB1}
\end{equation}
Thus, the exactness of SWKB guarantees that adding the generalized Langer-correction  term $\frac 14 \,\hbar^2 f_1'(x)$ to the potential $V_-(x,a)$ makes the WKB quantization condition exact for all conventional SIPs. 

We list these generalized corrections for all three classes in Table 1. 
For each specific SIP, the form of $f_1$ can be found in Appendix \ref{AppendixA}. We note that for the Class I potentials, \(f_1'=0\), and thus no correction is needed. We also note that for both the Coulomb and 3D-oscillator potentials, $f_1=-1/r$; in these two cases,  $\Delta V(r) = \frac 14 \,\hbar^2 f_1'(r)=\frac{\hbar^2}{4r^2}$ becomes the traditional Langer correction\cite{Kramers1926,Langer1937}.
\begin{table} [htb]
	\centering
	\begin{tabular}{l|l|l}
		Class&Form of $W$ &  Generalized Langer    Correction $(\hbar^2/4)f_1^\prime $\\ 
			\hline \hline 
			Class I& $f_2(x)+\alpha\, a$ & $0$\\
			Class II&$a f_1(x)+B/a$ & $\frac{\hbar^2}{4} (f_1^2-\lambda)$\\
			Class III& $a f_1(x)+f_2(x)$ & $\frac{\hbar^2}{4} \, \left( f_1^2 -\lambda\right) $\\
			\end{tabular}	\label{table:classes} 
			\caption{The generalized langer correction for all three classes.}
\end{table}

\section{SWKB for Extended Shape Invariant Superpotentials}

\label{sec.Extended}
In Section \ref{sec:SIPs}, we saw that additive shape invariant superpotentials belong to one of two classes: conventional and extended. In this section we will show that  extended superpotentials do not satisfy SWKB exactness by providing a  counterexample.

Extended superpotentials $W$ depend explicitly on $\hbar$. They can be written as 
\begin{equation}
	\label{kernel}
	W(x,a,\hbar)=W_0(x,a)+W_h(x,a,\hbar),
\end{equation}
where the ``kernel'' $W_0$ does not depend explicitly on $\hbar$, and $W_h$ includes all the explicit $\hbar$-dependence. For all known cases, the kernel \(W_0(x,a)\) is a conventional superpotential. The authors of Refs. \cite{Bougie2010,Bougie2012} showed that the energy spectrum of extended superpotentials is determined solely by $W_0$. Thus each known extended SIP is isospectral with its conventional counterpart.

Despite the exactness of SWKB for conventional SIPs, extended superpotentials do not necessarily satisfy the SWKB condition in spite of being additively shape invariant. 
This was proven numerically in Ref.~\cite{Bougie2018} by considering a particular example:

\begin{equation}
	W\left(x,\ell,\hbar\right)=\frac{1}{2}\,\omega x-\frac{\ell}{x}+\frac{2\omega x\hbar}{\omega x^2 + 2\ell - \hbar}-\frac{2 \omega x \hbar}{\omega x^2+2\ell+ \hbar}~.
	\label{12a}
\end{equation}
This extended superpotential \cite{Quesne2008} has the 3-D oscillator superpotential
\begin{equation} 
	W_0=\frac{1}{2}\,\omega x-\frac{\ell}{x}~
	\label{eq:W0}
\end{equation}
as its kernel, and the \(\hbar\)-dependent extension
\begin{equation}
	W_{h}=\frac{2\omega x \hbar}{\omega x^2 + 2\ell -\hbar}-\frac{2 \omega x \hbar}{\omega x^2+2\ell+\hbar}~.\label{eq:hbar1W}
\end{equation}
The extended superpotential \(W(x,\ell,\hbar)\) of Eq. (\ref{12a}) is additively shape invariant and isospectral with the 3-D oscillator. 

We introduce the dimensionless variable \(z = x\sqrt{\omega/\hbar}\) and the dimensionless parameter \(\lambda = \ell/\hbar\), to obtain
\begin{equation}
	\label{eq:A}
	W(z,\lambda,\hbar) = \sqrt{\hbar\omega} \left[
	\frac{z}{2}  - \frac {\lambda}{z}
	+ \frac{2 z}{z^2 + 2\lambda - 1}
	- \frac{2 z}{z^2 + 2\lambda + 1}
	\right]~.
\end{equation}
Thus, the SWKB integral  $I\equiv\int_{x_1}^{x_2}\sqrt{E_n-W^2(x)}~{\rm d}x$ becomes
\begin{equation}
	I = \hbar\int_{z_1}^{z_2} 
	\sqrt{\eta\left(z,\lambda\right)}\quad{\rm d}z,
	\label{eq:integral0}
\end{equation}
where $\eta(z,\lambda)$ is the dimensionless quantity:
\begin{equation}
	\label{eq:A1}
	\eta(z,\lambda)
	= 2n - 
	\left[
	\frac{z}{2}  - \frac {\lambda}{z}
	+ \frac{2 z}{z^2 + 2\lambda - 1}
	- \frac{2 z}{z^2 + 2\lambda + 1}
	\right]^2~.
\end{equation}
Thus, the SWKB integral reduces to Eq. (\ref{eq:integral0}) which is $\hbar$ times  a dimensionless integral. 
Numerical analysis proved \cite{Bougie2018} that the integral  is not equal to \(n\pi\), therefore the extended superpotential of Eq. (\ref{12a}) does not satisfy the SWKB condition. Consequently, extended superpotentials are not necessarily SWKB exact.

In Ref.~\cite{Nasuda2021}, the authors use numerical analysis to demonstrate that several other examples of extended potentials are similarly not SWKB exact; the authors initially claimed that their analysis also improves upon the treatment of $\hbar$ used in Ref.~\cite{Bougie2018}; however, the two approaches are equivalent, as demonstrated in Ref.~\cite{Bougie2023}.

\section{Conclusions}\label{sec:conclusions}

Exactly solvable potentials are relatively rare in quantum mechanics, so approximation methods such as WKB  continue to play an important role in analyzing quantum systems.
Solvable potentials are highly instructive in deepening our understanding of approximation methods and suggest new directions for future development. 
In this manuscript, we reviewed recent advances related to mathematical properties that ensure exact solutions of the WKB and related SWKB conditions for all conventional shape-invariant superpotentials.

\begin{itemize}
	\item The SWKB quantization condition has long been known to produce exact eigenvalues for conventional SIPs in their unbroken phase \cite{Comtet1985,Dutt1986,Eckhardt1986,Raghunathan1987,Dutt1991}. In  Sec.~\ref{sec:UBSWKB}, we reviewed recent scholarship \cite{Gangopadhyaya2020}  proving that this exactness is a consequence of the mathematical structure underlying  conventional shape-invariance. This result provides a key to the entire set of advances discussed in this manuscript.
	\item In the broken phase, the related BSWKB condition applies \cite{Inomata1994, Inomata1993a, Inomata1993b, Junker2019, Eckhardt1986}. In Sec~\ref{sec:BSWKB}, we reviewed recent work \cite{Gangopadhyaya2020} showing that BSWKB is exact for all conventional SIPs in which the relevant integral is properly defined. This exactness results from a discrete shape-invariance connecting the broken and unbroken phases, and thus follows from the exactness of the corresponding unbroken phase.
	\item The WKB condition is exact for the harmonic oscillator and Morse potentials, but not for the other conventional SIPs. However, adding the ``Langer correction,'' initially identified by Kramers \cite{Kramers1926}, to the Coulomb and 3D-oscillator potentials makes them exact as well\cite{Langer1937, Langer1949, Bailey1964, Froman1965}. In Sec~\ref{sec:langer}, we reviewed recent work that generalized the Langer correction to all conventional SIPs\cite{Gangopadhyaya2023}. This result derives directly from the SWKB exactness of these superpotentials.
	\item Extended SIPs are additive shape-invariant, however recent numerical analysis found extended SIPs that are not SWKB-exact \cite{Bougie2018, Nasuda2021, Bougie2023}. Therefore, the SWKB exactness of conventional SIPs does not arise from additive shape-invariance alone, but also assumes \(\hbar\)-independence of the corresponding superpotentials. We discussed the SWKB condition for extended SIPs in Sec.~\ref{sec.Extended}.	
\end{itemize}

Together, these recent advances demonstrate that the mathematical framework of conventional SIPs has several important consequences. First, it ensures the exactness of the SWKB condition for these SIPs in their unbroken-SUSY phase. It also guarantees their BSWKB-exactness and provides for a generalized Langer correction; these two results can be directly related to the SWKB-exactness. Finally, extended SIPs are not necessarily  SWKB exact, despite being additive shape-invariant. Thus, the form of conventional SIPs  has wide-ranging consequences for semiclassical approximations, giving these potentials an important status for further developing our understanding of semiclassical methods.

\appendix
\newpage
\section{The Complete List of Conventional SIPs}\label{AppendixA}	

As we have seen from Eq.~\ref{eq.PDE2solution}, all conventional SIPs must be of the form 	
\[W(x,a) = a f_1(x) + f_2(x)+u(a).\]
Throughout this paper, we have categorized these solutions into three classes:
\begin{itemize}
	\item Class I: $f_1(x)$ is a constant, so the $x$-dependence of $W$ is only through $f_2(x)$.
	\item Class II: $f_2(x)$ is a constant, so the $x$-dependence of $W$ is only through $a f_1(x)$.
	\item Class III: Neither $f_1(x)$ nor $f_2(x)$ are constant, so that both $a f_1(x)$ and $f_2(x)$ contain functional $x$-dependence.
\end{itemize}

In Table 
2, we list the complete list of conventional SIPs\footnote{The general form of the Morse superpotential is $W=-\alpha A -B e^{\alpha x}$. In the table we have chosen $\alpha=-1,B=1$ to match the literature.}, together with the type and class/subclass to which they belong.

\begin{landscape}
\begin{table}[ph]
{
				\begin{tabular}{l|l|l|l|l}
					\textbf{Type}&\textbf{Category}& \textbf{Superpotential $W$}&\textbf{Potentials $V_{\pm}$ }&\qquad \textbf{Name}\\ 
					\hline \hline 
					II &IA &$\frac12 \omega x$ & $\frac14 \,\omega ^2x^2 \pm \frac12 \, \hbar \omega$ &Harmonic Oscillator  \\
					&& (no $a$ dependence, $f_1=0$)&& \\
					\hline
					II&IB &$A-e^{-x}$&$A^2-\left( 2 A \mp \hbar \right) e^{-x}+e^{-2 x}$ &Morse\\
					&&$\left(a=-A,\; f_1 {\rm \;is \; a  \; constant}\right)$&& \\
					\hline
					II&IIA&$-\frac \ell r+ \frac{e^2}{2\ell}~$ & $-\frac{e^2}{r}+\frac{\ell(\ell \pm \hbar)}{r^2} +\frac{e^4}{4 l^2}$ & Coloumb\\
					&&$\left(a=\ell, \quad B = \frac{1}{2}e^2\, ,  \quad f_1=-\frac1r\right)$&& \\
					\hline
					I&IIB1&$-A\,\cot x - \frac BA$& $A \left(A\pm\hbar \right)  \csc ^2x +$ & Rosen-Morse  \\
					&& $\left(a=A, \quad f_1=-\cot x\right)$ & $2 B \cot x+\frac{B^2}{A^2}- A^2$& (Trigonometric)\\
					\hline
					I&IIB2&$ A\,\tanh x + \frac BA~$&
					$-A \left(A\mp\hbar \right)  \text{sech}^2x+$&Rosen-Morse \\
					&&$\left(a=-A,~B\to -B, \quad f_1=-\tanh x\right)$ &$2 B \tanh x+\frac{B^2}{A^2}+A^2$& (Hyperbolic) \\
					\hline
					I&IIB3&$-A\coth r + \frac BA~$& $A (A\pm \hbar)\, \text{csch}^2r-$   &Eckart\\
					&&$\left(a=A, \quad f_1=-\coth r\right)$&$2 B \coth r  + \frac{B^2}{A^2}$& \\
					\hline
					II&IIIA&$\frac12\, \omega r -\frac \ell r$& $\frac14\, \omega ^2\,r^2 +\frac{\ell(\ell\pm\hbar)}{r^2}-\left( \ell \mp  \frac{h }{2} \right) \omega$  & 3D-Oscillator\\
					&&$\left( a= \ell, \quad f_1=-\frac1r\right)$&&\\
					\hline
					I&IIIB1&$A\tan x - B\, {\rm sec\,}x$ & $\left(A (A\pm \hbar)+B^2\right)\sec^2x -$ & Scarf  \\	
					&&$\left(a=A, \quad f_1=\tan x\right)$&$B (2 A\pm\hbar) \tan x \sec x -A^2$& (Trigonometric)\\
					\hline
					I&IIIB2&$A\tanh x + B \, {\rm sech\,}x$  &
					$- \left(A(A \mp h)-B^2\right){\rm sech}^2x+$&Scarf \\
					&&$\left(a=-A, \quad f_1=-\tanh x\right)$&$B (2 A\mp h) \tanh x \,{\rm sech}\,x+A^2$& (Hyperbolic)\\
					\hline
					I&IIIB3& $A\coth r - B \, {\rm csch\,}r$ & $ \left(A(A\mp  \hbar)+B^2\right)\text{csch}^2r-$ &P\"oschl-Teller \\
					&&$\left(a=-A, \quad f_1=-\coth r\right)$&$B (2 A\pm \hbar) \coth r \,\text{csch}\,r+A^2$& (Hyperbolic) \\
					\hline 
				\end{tabular}\label{table:SIPs}	}
	\caption{A complete list of all conventional shape-invariant superpotentials. This table presents $W$ for each superpotential together with the corresponding potentials $V_{\pm}$, parameter $a$, and functional form of $f_1$. The form of $f_1$ excludes constant terms, which cannot effect the Langer correction and can be arbitrarily included in either $a f_1$ or $u(a)$.}
\end{table}
\end{landscape}

\newpage
\section{Interrelations between type-I and type-II superpotentials}\label{AppendixB}	

\begin{table}[htbp]
			{
		\begin{tabular}{l|l|l}  
			{\bf Type-I Superpotential}   & {\bf Projection }                     & {\bf Type-II Superpotential } \\ \hline \hline 
			{\bf 1) Scarf (Hyperbolic)}   & $P_{1a}:$	                          & {\bf a) Morse } 	          \\
			$W(x)=A\tanh\left( x + \beta\right)+B~\textrm{sech}\left( x + \beta\right)$       &                       
			$A\to A$                                                                  & 
			$W(x)=A- B\, e^{-x}$                                                                                       \\
			$-\infty<x<\infty, \;A>0 $ 	                                                      & 
			$B\to - B\, {e^\beta \over 2}$                                              & 
			$-\infty<x<\infty $                                                                                   \\
			$E_n=A^2-\left(A-n \hbar \right)^2$  &	$\beta\to \infty$             & $E_n=A^2-\left(A-n \hbar \right)^2$  \\ \hline
			{\bf 2) Generalized P\"oschl-Teller }  & $P_{2a}$:                    & {\bf a) Morse }               \\
			$W(r)=A\coth\left(\alpha r + \beta\right)-B~\textrm{cosech}\,\left(\alpha r + \beta\right)$         &
			$A \to A $ 	                                                              & 
			$W(x)=A-B\,e^{ -x}$                                                                                      \\
			{~~~}        & $B\to B\, {e^\beta \over 2}$, $\alpha \to 1$                   & $-\infty<x<\infty $           \\ 
			{~~~}        & $\beta\to \infty$, $r \to x$           & $E_n=A^2-\left(A-n\hbar \right)^2$  \\ \cline{2-3}
			{~~~}                         & $P_{2b}:$                             & {\bf b) 3-D Oscillator }      \\
			$-\frac{\beta}{\alpha} < r<\infty$			                                                      & 
			$A \to \left({\omega \over \alpha} - \frac{\alpha\ell}{2}\right)$         & 
			$W(r)=\frac{1}{2}\omega r -\frac{\ell}{r}$                                                            \\
			$E_n=A^2-\left(A-n\alpha\hbar \right)^2$					                                  &
			$B \to \left({\omega \over \alpha} +  \frac{\alpha\ell}{2}\right)$        & 
			$0<r<\infty$                                                                                          \\
			$A<B$       & $\beta\to 0$, $\alpha \to 0$            & $E_n=2n\omega\hbar $                \\ \hline
			{\bf 3)  Scarf (Trigonometric) }     & $P_{3b}:$                      & {\bf b) 3-D Oscillator}       \\
			$W(x)=A \tan\left(\alpha x\right) - B\sec\left(\alpha x\right)$		              &
			$A \to \left({\omega \over \alpha} + \frac{\alpha\ell}{2}\right)$         & 
			$W(r)=\frac{1}{2}\omega r -\frac{\ell}{r}  $                                                          \\
			$-{\pi \over 2\alpha}<x<{\pi \over 2\alpha}, 	~A>B$                             &
			$B \to \left({\omega \over \alpha} -  \frac{\alpha\ell}{2}\right)$        & 
			$0<r<\infty$                                                                                          \\
			$E_n=\left(A+n\alpha\hbar\right)^2-A^2$                                                & 
			$x\to r+{\pi \over 2\alpha}$, $\alpha \to 0\; $			          & 
			$E_n=2n\omega\hbar $                                                                                       \\ \hline
			{\bf 4) $  $Rosen-Morse I}    & $P_{4c}:$		                      &	{\bf c) Coulomb}		      \\
			$W(x)=-A\cot\left(\alpha x\right)-\frac{B}{A}$	                                  &
			$A\to \alpha \ell$					                                      & 
			$W(r)=\frac{e^2}{2\ell}	-\frac{\ell}{r}$	                                                          \\
			$ 0 < x <\frac{\pi}{\alpha} $	                              &
			$ B \to - {\alpha \over 2} e^2$		 				                      & 
			$0<r<\infty $ 	                                                                                      \\
			$E_n=- A^2+\left(A+n\alpha\hbar\right)^2+{B^2 \over A^2} -{B^2 \over {(A+n\alpha\hbar)}^2}$ &
			$\alpha \to 0$, $x\to r$	                                          & 
			$E_n={e^4 \over {4\hbar^2}}\left( {1\over l^2}-{1\over (n+l)^2}\right)$	                              \\ \hline
			{\bf 5) $  $Rosen-Morse II}      & {~~} ---                           & {~~} ---	                  \\
			$W(x)= A \tanh(x) + \frac BA$    & {~~}                               & {~~}                          \\
			$-\infty<x<\infty $, $B < A^2$   & {~~}                               & {~~}                          \\
			$E_n= A^2-\left(A-n\hbar\right)^2-\frac{B^2}{(A-n\hbar)^2} +\frac{B^2}{A^2}$ & {~~} &                 \\ \hline
			{\bf 6)  Eckart}              & $P_{6c}:$ 		                      & {\bf c) Coulomb}              \\
			$W(r)=-A\rm{coth}\left(\alpha r\right)+\frac{B}{A}	$	                          &
			$A\to \alpha \ell$						                                  & 
			$W(r)=\frac{e^2}{2\ell}-\frac{\ell}{r} $                                                              \\
			$0<r<\infty,\; B>A^2,\; A>0$ & $ B \to {\alpha \over 2} e^2$  & $0<r<\infty $                 \\
			$E_n= A^2-\left(A+n\alpha\hbar\right)^2+{B^2 \over A^2} -{B^2 \over {(A+n\alpha\hbar)}^2}$  &
			$\alpha\to 0$                                                             & 
			$E_n={e^4 \over {4\hbar^2}}\left( {1\over l^2}-{1\over (n+l)^2}\right)$                                  \\ \hline 
		\end{tabular}\label{table:reductions}}
\caption{Limiting procedures and redefinition of parameters relating type-I to type-II superpotentials. For projections, in each cell the order of operators should be carried out from top to bottom. Adapted with permission from Mallow, et al.\cite{Mallow2020} \copyright 2019 Elsevier B.V.}
\end{table}

\newpage
\section{List of Relevant Integrals}\label{AppendixC}	

\noindent
These integrals are listed in the Appendix of Ref.~\cite{Gangopadhyaya2023}. Many of these integrals previously appeared in Ref.~\cite{Hruska1997}. 
\begin{eqnarray}
	I_0(y_1,y_2) &\equiv& \int_{y_1}^{y_2} ~dy~
	\sqrt{\left( y_2-y\right)\left( y-y_1\right)}
	=\frac{\pi}{8}\left( y_2-y_1\right)^2~;~
	\nonumber \\	
	I_{1a}(y_1,y_2) &\equiv& \int_{y_1}^{y_2} \frac{~dy~
		\sqrt{\left( y_2-y\right)\left( y-y_1\right)}}{y}
	=\frac{\pi}{2}\left( y_1+y_2\right) - \pi\sqrt{y_1 \, y_2}~,~(0<y_1<y_2)~;~
	\nonumber \\	
	I_{1b}(y_1,y_2) &\equiv& \int_{y_1}^{y_2} \frac{~dy~
		\sqrt{\left( y_2-y\right)\left( y-y_1\right)}}{y}
	=\frac{\pi}{2}\left( y_1+y_2\right) + \pi\sqrt{y_1 \, y_2}~,~(y_1<y_2<0)~;~
	\nonumber \\	
	I_{2a}(y_1,y_2) &\equiv&  \int_{y_1}^{y_2}
	\frac{~dy~
		\sqrt{\left( y_2-y\right)\left( y-y_1\right)}}{y^2} =
	-\frac{\pi \left( {y_1}+{y_2} + 2 \sqrt{{y_1} {y_2}}
		\right)}{2 \sqrt{{y_1} {y_2}} }
	~,~(y_1<y_2<0)~;~
	\nonumber \\
	I_{2b}(y_1,y_2) &\equiv&  \int_{y_1}^{y_2}
	\frac{~dy~
		\sqrt{\left( y_2-y\right)\left( y-y_1\right)}}{y^2} =
	\frac{\pi \left( {y_1}+{y_2} - 2 \sqrt{{y_1} {y_2}}
		\right)}
	{2 \sqrt{{y_1} {y_2}} }
	~,~(0<y_1<y_2)~;~
	\nonumber \\	
	I_3(y_1,y_2) &\equiv& \int_{y_1}^{y_2} \frac{~dy~
		\sqrt{\left( y_2-y\right)\left( y-y_1\right)}}{1+y^2}
	=\frac{\pi}{\sqrt{2}}\left[ \sqrt{1+y_1^2}  \sqrt{1+y_2^2}  -{y_1 \, y_2}+1                  \right]^{1/2} - \pi~;~
	\nonumber \\	
	I_4(y_1,y_2) &\equiv& \int_{y_1}^{y_2} \frac{~dy~
		\sqrt{\left( y_2-y\right)\left( y-y_1\right)}}{1-y^2} =\frac{\pi}{2}\left[2- \sqrt {\left( 1-y_1\right) \left( 1-y_2\right)}
	- \sqrt {\left( 1+y_1\right) \left( 1+y_2\right)}
	\right] ~, \nonumber \\
	&&\qquad \qquad {\rm where}~	(-1<y_1<y_2<1)~;~
	\nonumber \\
	I_{5a}(y_1,y_2) &\equiv& \int_{y_1}^{y_2} \frac{~dy~
		\sqrt{\left( y_2-y\right)\left( y-y_1\right)}}{y^2-1} =\frac{\pi}{2}\left[\sqrt {\left( y_1+1\right) \left( y_2+1\right)}- \sqrt {\left( y_1-1\right) \left( y_2-1\right)} -2
	\right] ~, \nonumber \\
	&&\qquad \qquad{\rm where}~	(1<y_1<y_2)~;~ \nonumber \\
	I_{5b}(y_1,y_2) &\equiv& \int_{y_1}^{y_2} \frac{~dy~
		\sqrt{\left( y_2-y\right)\left( y-y_1\right)}}{y^2-1} =
	\frac{\pi}{2} \left[\sqrt{(y_1-1) (y_2-1)}-\sqrt{({y_1}+1) (y_2+1)}-2\right]
	~, \nonumber \\
	&&\qquad \qquad{\rm where}~	(y_1<y_2<-1)~.
	\nonumber \\	
\nonumber\end{eqnarray}

\end{document}